\documentclass[aps,prd,preprint,showpacs,nofootinbib]{revtex4}

\usepackage[dvips]{epsfig}
\usepackage{amsfonts,amssymb,amsmath}
\usepackage{amsfonts,amssymb,amsmath}

\newcommand{\be}{\begin{equation}}
\newcommand{\ee}{\end{equation}}
\newcommand{\ben}{\begin{displaymath}}
\newcommand{\een}{\end{displaymath}}
\newcommand{\bea}{\begin{eqnarray}}
\newcommand{\eea}{\end{eqnarray}}
\newcommand{\bean}{\begin{eqnarray*}}
\newcommand{\eean}{\end{eqnarray*}}

\def\l {\lambda}
\def\a {\alpha}

\def\g {\gamma}

\newcommand{\bPp}{\bar{P}_-}

\newcommand{\eE}{E}
\newcommand{\eK}{K}

\newcommand{\ads}[1]{\mbox{${AdS}_{#1}$}}

\newcommand{\tr}{\mbox{Tr}}

\newcommand{\beq}{\begin{equation}}
\newcommand{\eeq}{\end{equation}}
\newcommand{\beqr}{\begin{displaymath}}
\newcommand{\eeqr}{\end{displaymath}}
\newcommand{\beqa}{\begin{eqnarray}}
\newcommand{\eeqa}{\end{eqnarray}}
\newcommand{\beqar}{\begin{eqnarray*}}
\newcommand{\eeqar}{\end{eqnarray*}}

\newcommand{\cN}{{\cal N}}

\newcommand{\cO}{{\cal O}}
\newcommand{\cL}{{\cal L}}

\newcommand{\bi}{\ensuremath{\bar{\imath}}}

\newcommand{\N}[1]{\ensuremath{\cN=#1}}

\def\S{{\mathcal S} }
\def\J{{\mathcal J} }

\def \ov {\over}

\def \foot {\footnote}
\def \bi{\bibitem}

\def \tr {{\rm tr}}
\def \ha {{1 \over 2}}
\def \ci{\cite}

\def\S{{\mathcal S} }

\def \la{\label}
\def \sql {\sqrt{\lambda}}

\def \del {\partial}
\def\be{\begin{eqnarray}}
\def\ee{\end{eqnarray}}
\newcommand{\rf}[1]{(\ref{#1})}

\def \E {{\cal E}}
\def \eE {{\rm E}}
\def \eK {{\rm K}}
\def \eR {{\rm R}}

\begin{document}

%\vspace{ -3cm}
\rightline{Imperial-TP-AT-2008-5}

%\vspace{ 1cm}

\title{Spiky strings in $AdS_3 \times S^1$  and their $AdS-pp$-wave limits}
\author{R. Ishizeki}
\email{rishizek@purdue.edu}
\author{M. Kruczenski}
\email{markru@purdue.edu}
\author{A. Tirziu}
\email{atirziu@purdue.edu}
\affiliation{ Department of Physics, Purdue University,
525 Northwestern Ave., W. Lafayette, IN 47907-2036, USA}
\author{A.A. Tseytlin}
\email{tseytlin@imperial.ac.uk}
\affiliation{Blackett Laboratory, Imperial College,
London SW7 2AZ, U.K. \footnote{Also at
 Lebedev  Institute, Moscow.} }
\date{\today}

\begin{abstract}
\baselineskip 13pt
\

 We study a class of classical solutions  for closed  strings moving in
 $AdS_3\times S^1$ $\subset$ $AdS_5\times S^5$  with
 energy $E$ and spin $S$ in  $AdS_3$
and  angular momentum $J$ and winding $m$ in
 $S^1$. They have rigid shape  with $n$ spikes in $AdS_3$. We find that
 when $J$ or $m$ are non-zero, the spikes do not end in cusps.
We   consider in detail a special large $n$
 limit in which
$S\sim n^2, \ J\sim n$, i.e.  $S \gg J \gg 1$,
with
$E+S\ov n^2$, $E-S\ov n$, $J\ov n$ , $m\ov n$  staying   finite.
 In that limit the spiky spinning string approaches the boundary of $AdS_5$.
We show that the  corresponding  solution
 can be interpreted as describing
 a  periodic-spike string moving  in  $AdS_3$--pp-wave $\times S^1$
background.
 The resulting expression for the string energy
 should represent   a strong-coupling  prediction for
  anomalous dimension  of a class of dual gauge theory states in
 a particular thermodynamic limit of the $SL(2,R)$ spin chain.
\end{abstract}
%\vskip -4cm

%\pacs{11.25.-w,11.25.Tq}

%\keywords{AdS/CFT, string theory, anomalous dimensions}

%\preprint{}
%\vskip -4cm

\maketitle

%\fi

\section{Introduction}
\baselineskip 13pt

 Remarkable recent progress in understanding the spectrum of $AdS_5 \times S^5$ superstring theory
 was initiated, in particular,   by
  the study of various classical string
solutions. In particular,
the energy of the well-known folded spinning string solution in $AdS_3$  \cite{gkp}
describes the  dimension of
twist two gauge theory operators such as $\tr(\Phi \nabla_{+}^S \Phi)$ in the limit of large spin $S$.
The folded string solution was  generalized
to $AdS_5 \times S^1$ \cite{ft1} and quantum corrections to its energy
were computed \cite{ft1,ftt,rtt,rt2}.
%Although these computations are done at strong coupling,
These  and similar  results  on the string theory side aided  and tested
%were matched  by those
%the same results,
%in the large $S$ limit,
% obtained (for large $S$ and strong coupling) from the
 %from the field theory side
 the construction of all-loop asymptotic
 Bethe ansatz  for  anomalous dimensions of the dual gauge theory operators
(see,  e.g., \cite{bes, bbks, bkk,grom}).

The closed folded string solution \cite{gkp} was generalized  to  the case of
 $n$ spikes in $AdS_3$ in    \cite{spiky,krt1}. The corresponding gauge theory states  were argued
  \cite{spiky} to represent, in particular,
 a  subclass of higher twist operators in the  $SL(2)$ sector of gauge theory  \ci{bgk}.
 % where solutions with $n$ spikes were found in $AdS_3$.
In the large $S$ limit  the spikes approach
 the boundary of $AdS_3$. It was shown in \cite{ppwave} that the motion
of  spikes in this limit can be  described by a string  solution in an $AdS_5-pp$-wave
metric\footnote{Locally this space is still \ads{5}.}
%(here $x_{\pm}=\frac{1}{\sqrt{2}}(x\pm t)$)
\beq
 ds^2 = \frac{1}{z^2}\Big[ 2 dx_+ dx_-  - \mu^2 (z^2 + x_i^2)  dx_+^2 + dx_i dx_i   + dz^2 \Big] \ , \ \ \ \ \ \
\ \ \ \ \ i=1,2
\label{metric_xp}
\eeq
 The metric (\ref{metric_xp}) is obtained by zooming at  the near-boundary region of $AdS_5$ while
  at the same time moving close to the speed of light in an
angular direction. This  limit  thus appears to be
 relevant for the  study of strings  and hence  dual gauge theory operators  with large spin.
 String theory in an $AdS_5-pp$-wave $\times$
$S^5$ space is dual to $\N{4}$ SYM in a 4-dimensional   pp-wave  background with (conformally flat) metric
\beq\la{jk}
ds_{ft}^2 =  2 dx_+ dx_-  - \mu^2 x_i^2  dx_+^2 + dx_i dx_i
\eeq
Indeed, this  is the boundary metric of the space (\ref{metric_xp}).
In \cite{ikt} several solutions for strings moving in the  $AdS_3-pp$-wave space
were found. Those strings were ending at the boundary and therefore were dual
to various Wilson loops in  the field theory  in the boundary $pp$-wave background  \rf{jk}.

\

In this paper we consider new solutions were the  string
is entirely in the bulk and therefore should be dual to particular
states in the gauge  theory.
Interestingly,  such strings  do not move in direction $z$. In usual $AdS_5$ space in Poincare coordinates this is not possible for extended strings
since the curvature of the metric pushes the string towards small $z$. Here this effect is compensated
 by  the  extra term in the metric
\rf{metric_xp} proportional to $dx_+^2$.
 In fact,  the solutions we shall find can be seen as limits of the spiky string solution when
  the spikes  approach the boundary
while the number of spikes grows to infinity.
 As we shall discuss below, the relevant limit  that one can  take is
$n\rightarrow\infty$ keeping
 %$\bPp=
 $\frac{E+S}{n^2}$  and
$\bar{\g} = \frac{E-S}{n}$ fixed:
there is an infinite number of spikes each of which contributes
a finite amount $\bar{\g}$ to the anomalous dimension $\g=n \bar{\g}=E-S$.

In \cite{spiky} it was argued  that the spiky string should correspond
to an operator of the type
\beq\la{opa}
\cO = \tr\left( \nabla_+^{\frac{S}{n}}\Phi\ \nabla_+^{\frac{S}{n}}\Phi\ldots \nabla_+^{\frac{S}{n}}\Phi\  \right)
\eeq
 Such operators can be described by a spin chain  with a  number of sites $n$  being the
  same as the number of  spikes. This  correspondence  was recently
emphasized and extended  in  \cite{Dorey1} based also on earlier work of
\ci{bgk,kz}.
%Thus, what we find is that, the above described limit, should be a
This suggests that   the above large $n$ limit
should be a meaningful thermodynamic limit of such spin chain, describing
a  strong-coupling asymptotics  of the
corresponding anomalous dimension.
% We then predict the value of the
%function $\bar{\Delta} = f(\lambda, \bPp)$ in limit of string coupling,
%$\lambda\gg 1$.

\

More precisely, the operators  in the $SL(2)$ sector of planar $\cal N$=4 SYM theory are built out of
$J$ powers  of  complex scalar $\Phi$ and   $S$ powers   of  light-like  covariant derivative  $\nabla_+$,
 symbolically,  $\tr (\nabla_+^{S}\Phi^J)$. The eigen-states of the dilatation operator or spin chain Hamiltonian
are labeled  in addition to $S$ and $J$  by other quantum numbers  corresponding, e.g., to the
number of spikes $n$ or  $S^1\subset S^5$ winding   number $m$ on the dual string theory side.
Their  scaling dimension  may then be written as
\be \la{dim}
E= S + J   +  \gamma(S, J,  m, n; \lambda) \ ,
\ee
where $\l$ is 't Hooft coupling.
We shall assume  that the spin chain length $J$   is  large enough   so that one  can ignore
the wrapping contributions  \ci{wrap},
i.e. that $\g$  should have the  asymptotic Bethe ansatz \ci{bes} description  for all   values of $\l$.
%As we are interested in the large spin limit,
On the  perturbative gauge theory side one may consider the
 limit  of large $S$ (and large $J$)
at fixed $\lambda$, e.g., at each order in expansion in
$\l  < 1$.
 To describe the corresponding states in terms of semiclassical strings  one is to consider
 first $\l \gg 1$   with fixed $\S\equiv { S \ov \sql},\ \J\equiv { J \ov \sql}$  and
  then take $\S$ large  order by order in $1 \ov \sql$ expansion.
  The two expansions are not a priori the same   and
 may require a certain  resummation in order to match.

\

Here we   propose  to consider the following  special case  of the
$S \gg J \gg 1$ limit on the gauge-theory  side  \rf{dim}:
\be\la{lii}
 n \gg 1, \ \ \ \ \ \
{\rm  with} \ \ \ \ \
{E+S\ov n^2} = A, \ \ \ {E-S\ov n}= B , \ \  \ {J\ov n} = K , \ \ \ {m\ov n} =k  \ \ \    { \rm    fixed}
\ee
This limit is to be taken at fixed  $\l$, i.e. the fixed  ratios   may be
 functions of $\l$. \foot{Since $S,J,m,n$ should
be   integers it  appears that  only $B$
 can be a nontrivial function of $\l$.}
Equivalently, we assume that  for $n \to \infty$
\bea \la{asu}
&& E= \ha  A n^2   + \ha  B n   + O(n^0) \ , \ \ \ \ \ \  \ \ \ \ \
S= \ha  A n^2   - \ha  B n   + O(n^0) \ , \nonumber  \\
&& J=   K   n  + O(n^0) \ , \ \ \ \ \ \  \ \ \ \ \
m=   k   n  + O(n^0) \ . \ee
Then $E \sim n^2, \  S \sim  n^2, \ J \sim n $   so that  $ S   \sim n J$, i.e. $S \gg   J\gg 1 $.
The operator \rf{opa}  represents a  particular state that may be relevant in such limit having
$J=n$. More generally, one may consider
\beq\la{wpa}
\cO = \tr\left( \nabla_+^{\frac{S}{J}}\Phi\ \nabla_+^{\frac{S}{J}}\Phi\ldots \nabla_+^{\frac{S}{J}}\Phi\  \right)
\sim  \tr\left( \nabla_+^{ a  n }\Phi\ \nabla_+^{a n }\Phi\ldots \nabla_+^{a n }\Phi\  \right)
\eeq
with $a= { A \ov 2K}$.

To define a similar limit on the semiclassical string theory side
we need again to remember that to have  a consistent $\a' \sim { 1 \ov \sql}$
expansion we are to take  $\l \gg 1$ first  with all the  parameters  characterising  a
classical string solution like  $\S={ S \ov \sql},\ \J={ J \ov \sql}, \  m $  {\it and} $n$
being fixed, so that the string energy admits the expansion
\be \la{ew}
E = \sql \E_0 ( \S,\J, m, n) + \E_1 ( \S,\J, m, n) +   {1 \ov \sql}  \E_2 ( \S,\J, m, n)  + ...  \ . \ee
Then  the analog  of the limit \rf{lii} in the
perturbative string theory  expansion
is proposed to be
\be\la{kii}
 n \gg 1, \ \ \ \ \ \
{\rm  with} \ \ \ \ \
{\E+\S\ov n^2} = {\cal A} , \ \ \ {\E-\S\ov n}= {\cal B}  , \ \  \ {\J\ov n} = {\cal K} , \ \ \ {m\ov n} =k   \ \ \  { \rm    fixed}
\ee
where  $\E= { E \ov \sql}$.

 Below we shall consider  only the classical string solutions
for which the fixed parameters  in \rf{kii} will  not depend on $\sql$,
i.e. we will be interested in a  particular
 scaling limit in the space of semiclassical parameters.\foot{In principle, one may
consider a  more general  limit in  which   the fixed
 parameters  may be
given by  series in inverse string tension like  $c_0 + {c_1 \ov \sql}  + {c_1 \ov (\sql)^2} + ... $.
That would  correspond to a certain resummation  of string perturbative expansion.}
In this limit  $\E \sim n^2, \ \ \S \sim  n^2,\ \ \J \sim n $   so
that    $ S   \sim  \sql n^2 \gg n^2 , \ \     J \sim \sql  n \gg n  $.
This limit  is obviously different   from the one \rf{lii}
on the gauge theory (spin chain) side  but as with other large spin limits
the two may happen to be closely connected in certain  special cases (like leading terms in
large spin expansion).
%They should also be related  in the following sense:
Having  found the exact expression for the  dimension  $E$  in \rf{dim}
for all  values of $S,J,m,n$ and $\l$  one should be able to   consider
the large $n$ limit either  as   in \rf{lii} or as in \rf{kii}.  For certain terms (like familiar $\ln S$ terms)
the predictions of the two limits may differ only by  interpolating functions of $\l$,  but in general
to  connect the expressions found  in  the
two  limits should  require a resummation of the corresponding expansions.

%One implication of the above discussion  is
 Let us mention   that in the semiclassical string theory limit  that
we shall consider   below the  states with  R-charge   $J \sim n$
 like those in \rf{lii}  will not
be distinguishable  from states with $J=0$ :  to have a non-zero  semiclassical spin one would need
to consider  the states with $J \sim \sql  n\gg n  $.  In other words,  the  spiky strings moving only
in $AdS_5$ may still be thought of  as  corresponding    to $SL(2)$ spin chain operators
 like \rf{opa}.

\

%%%%%%%%%%%%%%%%%%%%%%%%%%%%%%%%%%%%%%%%%%%%%%%%%%%%%%%%%%%%%%%%%%%%%%%%%%%%%%%%%%%%%%%%%%%%%%%%%%%%%%%%%%%%%%%%%%%%%%%%%%%%

Below we shall also
 extend the discussion of the spiky strings in \ci{spiky}
 % to include an angular momentum $J$ in the $S^5$.
 %, namely $R$-charge from the field theory perspective.
 %For that purpose we first generalize the spiky string
to include the angular momentum $J$ (and winding $m$)
 in a maximal   circle $S^1$ in
 % (understood as a maximum circle of the
$S^5$. To obtain  the solution it turns out to be
 convenient to use the
conformal gauge  as in \cite{krt1,Jevicki}.  The resulting
solutions are closely related
to those discussed  in \cite{hosv}.
 We shall  find  that the shape of the string in the $\ads{3}$ space  is very similar to the original
 spiky string with $\J=m=0$.
 A careful analysis shows, however, that as long as
 $\J$ or $m$ are  non-zero
  the spikes are rounded,   namely,  they do not end in cusps.
%  It is interesting to notice this difference, however,
% it is not crucial in any respect. These new solutions are interesting in themselves.

The introduction of $\J$ allows for the possibility of taking the large $\J$
or fast-string limit as in  \cite{bmn,ft1,ft2,k,krt2}.
 The leading-order term  in the  string energy  is then described \ci{k,krt2}  by an effective
  $SL(2,R)$  Landau-Lifshitz (LL)   model \cite{st,ptt}  that happens to  capture  both
the fast-moving string limit and the corresponding leading-order semiclassical
 dynamics of the spin chain on  the perturbative field theory side.
%In that case perfect agreement is found
%between both sides of the AdS/CFT correspondence
We shall  show how   to find the spiky-like  solutions  directly in the  LL model
 which can then
be interpreted either
 as fast-moving strings  or as coherent superpositions of field theory operators.

 Another limit
 of interest is when $\frac{\E-\S}{n}$ is taken to be
much larger that $\frac{\J}{n}$. In that case we shall recover the familiar
logarithmic scaling of the anomalous dimension.

 Finally, we shall also
 %we go back to our initial intention which was to
  consider the $AdS-pp$-wave limit  which for  non-zero $\J$   corresponds to taking $n \to \infty$  with the ratios in
   \rf{kii} being fixed.
%\beq
%n\rightarrow\infty,\ \ \ \mathrm{with}\ \ \bPp=\frac{E+S}{n^2} , \ \ \ \ \bar{\Delta}=\frac{E-S}{n},\ \bar{J}=\frac{J}{n},\ \bar{m}=\frac{m}{n}\ \ \mathrm{fixed}.
%\eeq
 Here   each spike contributes a finite amount to the anomalous dimension $\E-\S-\J$, the
 spin $\J$ and the winding $m$. In this limit, which  should correspond, as discussed above,
to a  particular  thermodynamic limit on the spin chain side, we compute the
classical  string energy or $\E-\S-\J\ov n $
%or $\bar \g = { \g \ov n} $
 as a function of fixed parameters
%$\bPp=
 $ { \E + \S \ov n^2},\  {\J \ov n} , { m \ov n}$.
% $\bar{\Delta} = f(\bPp,\bar{J},\bar{m})$.
\foot{More precisely, this  function is defined
implicitly by computing %$\bar{\Delta}$, $\bPp$, $\bar{J}$ and $\bar{m}$
these ratios in terms of the
three independent parameters,  allowing in principle to obtain any of
the four quantities in terms of the other three.}
This function should represent  a  strong-coupling prediction
%as viewed from the gauge theory side
%Again we have a prediction for the
for the thermodynamic limit of the corresponding $SL(2)$ spin chain.
 It would be
very interesting if the methods generalizing those used for the scaling function
 \cite{bes, bkk} can be used to reproduce this prediction from the Bethe
  ansatz.\foot{As was already mentioned,
 in the limit we consider $J \to \infty$  so
 % applied to obtain such non-trivial function.
% An important point is that, since we consider an infinite chain,
that the  wrapping contributions should be absent.}
%play any role in the result.

\

This rest of this paper is organized as follows.  In section 2 we shall review the original
spiky string solution with $\J=0$ and show that it admits
a consistent  large $n$ limit as defined in \rf{kii}.
We shall then demonstrate that  the same  expression for its energy can be found by
considering an infinite  rigid  string  with periodic spikes  in the $AdS_3-pp$-wave background \rf{metric_xp}.

As an aside, in section 3 we shall study a straight  string in $AdS$-- pp-wave  background
and show that expanding its energy at large $S$ one is able  to reproduce certain terms in the
large  $S$ expansion
of the folded (2-spike)   string  in $AdS_5$  which is an indication
 of the utility of the pp-wave  picture.

In section 4 we shall describe in detail the construction of the  generalization of the $n$-spike
 solution to the presence of
  non-zero classical $S^5$  angular momentum $J$ and winding  $m$.
  We shall follow  \ci{krt1} and use a generalized rigid string ansatz in  the conformal gauge.
  We shall show that a non-zero $J$ or $m$   ``rounds-up'' the spikes and find the (implicit) expression
  for the energy as a function of the semiclassical parameters $\S, \J, n, m $.
  Then in section 5 we shall  consider the  three special  cases: (a) the  $\J=m=0$ case  when the solution reduces to the
  original spiky string   in $AdS_3$; (b) the fast-string limit with $\J \gg 1, \   {\S \ov \J}=$fixed,
  which should be reproduced by the   Landau-Lifshitz  model; (c) the large $n$ limit \rf{kii}
  where  $\S \gg \J \gg 1$ and
  which should also admit a description in terms of a
   rigid string in $AdS$--pp-wave background.

  In section 6  we shall elaborate on the  connection to the  Landau-Lifshitz  model
  by presenting the corresponding analogs of the spiky string solution in several different limits.
  In section 7 we shall demonstrate   how to construct the the generalization of the  periodic spike solution
  in $AdS$--pp-wave background from section 2 to the presence  of rotation in an extra $S^1 \subset S^5$
  and discuss its connection to the  large $n$
 limit of the solution found in  section 4.
Appendix contains a list of some useful integrals.

\section{Large $n$
 limit of spiky string  as periodic spike solution in  $AdS$-- pp-wave  background}

 In this section we consider a particular limit of the spiky string \cite{spiky} which corresponds to taking the number of spikes $n$ to infinity
keeping $(\E+\S)/n^2$ and $(\E-\S)/n$ fixed.
%, where $E$ is the energy and $S$ is the spin.
%Notice the different scaling with $n$ of $E\pm S$.
 It turns out that such limit can be  also
described by a particular spiky  solution for a
 string moving in an $AdS$--pp-wave background.
This follows
from the fact that, at the level of  the string solution, this  limit
is the same as the one shown in \cite{ppwave} to lead to an
$AdS$--pp-wave metric.

\subsection{Limit of the spiky string solution}

 Following \cite{spiky} we consider a rigid string rotating around its center of mass
 in the $AdS_3$ metric
\beq\la{mam}
ds^2 = -\cosh^2\!\!\rho\, dt^2 +d\rho^2 + \sinh^2\!\!\rho\, d\theta^2
\eeq
and described by  the ansatz
\beq
 t = \kappa \tau, \ \ \ \theta=\omega\tau + \sigma, \ \ \ \rho=\rho(\sigma).
\eeq
 Then $\rho$ satisfies
% $v=\frac{1}{\cosh2\rho}$
 %which then satisfies
\footnote{In \cite{spiky}
a variable $u=\cosh 2\rho$ was used. Here we find it more convenient
 to use, instead, $v=1/u$.}
\beq
 \frac{dv}{d\sigma} =
(1-v^2) \sqrt{\frac{1-v_1}{1-v_0^2}} \sqrt{\frac{v_0^2-v^2}{v(v-v_1)}} \ , \ \ \ \ \ \
v\equiv \frac{1}{\cosh2\rho}
\eeq
where $v_{0}$ and $v_{1}$ determine the positions of the spikes
and the valleys, namely the maximal and the  minimal values of $\rho$ (or $v$).
 It is then straightforward to compute
\beqa
 \Delta \theta &=& \frac{2\pi}{n} = 2\sqrt{\frac{1-v_0^2}
{1-v_1}} \int_{v_1}^{v_0} \frac{dv}{1-v^2} \sqrt{\frac{v(v-v_1)}{v_0^2-v^2}} \\
 \frac{\S}{n} &=& \frac{\sqrt{1+v_1}}{4\pi v_0} \int_{v_1}^{v_0} \frac{dv}{v(1+v)} \sqrt{\frac{v_0^2-v^2}{v(v-v_1)}} \\
 \frac{\E}{n} &=& \frac{1-v_0^2}{2\pi v_0 \sqrt{1-v_1}}  \int_{v_1}^{v_0} \frac{dv}{1-v^2} \sqrt{\frac{v(v-v_1)}{v_0^2-v^2}}
       +  \frac{1-v_1}{4\pi v_0}  \int_{v_1}^{v_0}
 \frac{dv}{v(1-v)} \sqrt{\frac{v_0^2-v^2}{v(v-v_1)}}
\eeqa
 Here $n$ is the number of spikes,
$\Delta \theta$ is the angular distance between spikes, and  $E= 2\pi T \S$ and $S=2\pi T\S$
 are the energy and spin respectively ($T= { \sqrt \lambda \ov 2\pi}$
is the string tension).
 The resulting expressions
 can be written in terms of elliptic functions as in \cite{spiky} but the above integral representations
are
more convenient for taking the limit we are interested in here.

 Indeed, let us rescale
\beq
 v_0 \rightarrow \epsilon^2 v_0 , \ \ \ \ \ \ \ \ \ v_1 \rightarrow \epsilon^2 v_1\ , \ \ \ \ \ \ \ \
\epsilon\rightarrow 0\ .
\eeq
%and take It immediately
Then it follows from the above relations  that
\beq
 n \sim \frac{1}{\epsilon^2} ,\ \ \ \ \  \
 \ \ \frac{\E+\S}{n} \sim \frac{1}{\epsilon^2} ,\ \ \ \  \ \ \ \frac{\E-\S}{n} \sim 1
\eeq
 Thus we can compute the following
 finite quantities
\beqa
 \frac{\E+\S}{n^2} &\rightarrow& \frac{1}{2\pi^2}
 \int_{v_1}^{v_0}  \frac{dv}{v} \sqrt{\frac{v_0^2-v^2}{v(v-v_1)}}\
                                             \int_{v_1}^{v_0}  \frac{dv'}{v_0} \sqrt{\frac{v'(v'
-v_1)}{v_0^2-v'^2}} \\
 \frac{\E-\S}{n}   &\rightarrow&  \int_{v_1}^{v_0}  \frac{dv}{2\pi v_0} \sqrt{\frac{v(v-v_1)}{v_0^2-v^2}} +
                                  \int_{v_1}^{v_0}  \frac{dv}{2\pi v} \left(1-\frac{v_1}{2v}\right)\sqrt{\frac{v_0^2-v^2}{v(v-v_1)}}
\eeqa
 which are clearly invariant under the above rescaling:
 they depend only on the ratio
\be  b\equiv \frac{v_1}{v_0}  \ee
 This can be made explicit by  writing  them
in terms of the elliptic functions:
\beqa
 \bPp &\equiv & \frac{E+S}{n^2} = \frac{2T}{\pi b(1+b)} \Big[(1+b)\eE(p) - (2+b) \eK(p) -b^2
\Pi(1-b,p) \Big]^2 \nonumber \\
 \bar{\gamma} &\equiv & \frac{E-S}{n}= \frac{T}{\sqrt{1+b}}
\Big[ (1+b)\eE(p) - (2+b) \eK(p)
 + b^2 \Pi(1-b,p)\Big]
\label{res1}
\eeqa
 where we explicitly included the factor of string tension $T$  and defined
\beq
p = \sqrt{\frac{1-b}{1+b}}
\eeq
 $\eE(p)$, $\eK(p)$, $\Pi(1-b,p)$ are the standard elliptic functions
(defined  as  in \cite{Grad}).
 Eliminating $b$ one
finds $\bar{\gamma}$ as a function of $\bPp$.

As was already mentioned  in the Introduction,
  in \cite{spiky} it was argued that the spiky string should
be dual to  the operator \rf{opa} from the $SL(2)$ sector of gauge theory
 (with $J=n$ which is
not distinguishable from zero  at the classical string theory level).
%\beq
%\cO = \tr\left(\nabla_+^{\frac{S}{n}}\Phi\ \nabla_+^{\frac{S}{n}}\Phi\ldots \nabla_+^{\frac{S}{n}}\Phi\  \right)
%\eeq
%where $n$ is the number of $\Phi$ operators or equivalently the number of spikes.
%Such operator can be described by an $SL(2,R)$
%spin chain.
 The  large $n$ limit we just discussed should correspond to
a particular thermodynamic limit of the spin chain.
It would thus  be interesting    to reproduce the  above strong-coupling  expression for $\g$
%in the strong-coupling limit of the spin chain
% be certainly interesting to
%understand the result form a field theoretical point of view.
on the spin chain side.

\subsection{Periodic spike solution in  AdS--pp-wave background}
\label{pp-wave1}

 Let us  now show that exactly the same result (\ref{res1})
 can be obtained by considering a string moving in an $AdS$--pp-wave  \rf{metric_xp}. This space arises
as a particular limit of $AdS_5$ when one zooms in  near the boundary
while at the same time moving close to
the speed of light along the  angular
direction. This suggests
 that the $AdS$--pp-wave metric captures
all the information necessary to understand such thermodynamic limit.
This is considerably more than what
 was argued for  in \cite{ppwave} since there only the
leading $\ln \bPp$ dependence of $E-S$ was considered.

\subsubsection{Rigid strings in an AdS--pp-wave}

 As a preparation, let us start  by presenting a generic
 description of the relevant class of solutions in the metric \rf{metric_xp}, i.e.
%The space-time metric in which the string moves is
\beq
ds^2 = \frac{1}{z^2}\left[2 dx_+ dx_- - \mu^2 (z^2+x_i^2) dx_+^2 + dx_i dx_i + dz^2 \right]
\ , \ \ \ \ \ \ \ \ \ x_\pm = \frac{x\pm t}{\sqrt{2}}
\eeq
 %The only restriction is that we now consider strings such that
We shall assume that  $x_i=0$, i.e. consider strings
 that move in the subspace spanned by $x_{\pm}$,\ $z$.
To respect the symmetries of the problem we shall make a partial gauge choice by taking
\beq
x_+=\tau.
\eeq
 In this gauge the string action becomes
\beq
 I = \int {d\sigma d\tau}\ \cL\ = - T \int \frac{d\sigma d\tau}{z^2} \sqrt{x'_-{}^2+2x'_-\dot{z}z'-2\dot{x}_-z'{}^2+\mu^2z^2z'{}^2}
\eeq
 where dot indicate derivative with respect to $\tau$ and prime -- with respect to $\sigma$.
The equation of motion for $x_-$ is
\beq
 \partial_\sigma\left(\frac{x'_-+\dot{z}z'}{z^2F}\right) -\partial_\tau\left(\frac{z'{}^2}{z^2F}\right)=0
\eeq
and the one for $z$ is
\beq
\partial_\sigma\left(\frac{x'_-\dot{z}-2\dot{x}_-z'+\mu^2z^2z'}{z^2F}\right)+\partial_\tau\left(\frac{z'x'_-}{z^2 F}\right)
 =-\frac{2}{z^3}F + \frac{\mu^2z'{}^2}{zF}
\eeq
where
\beq
F = \sqrt{x'_-{}^2+2x'_-\dot{z}z'-2\dot{x}_-z'{}^2+\mu^2z^2z'{}^2}
\eeq
The conserved momenta are\foot{To find the expression for  $P_+$ it is convenient to assume
that  the gauge is  $x_+=\kappa \tau$ and compute
 $P_+=\int d\sigma \frac{\partial \cL}{\partial \kappa}$ before
setting $\kappa=1$.}
\beqa
 P_+ &=& \int d\sigma \frac{\partial \cL}{\partial\dot{x}_+}
                  = - T \int \frac{d\sigma}{z^2F} \left(x'_-{}^2+\dot{z}z'x'_--\dot{x}_-z'{}^2+\mu^2z^2z'{}^2\right)\nonumber\\
 P_- &=& \int d\sigma \frac{\partial \cL}{\partial \dot{x}_-} = T \int \frac{d\sigma}{z^2} \frac{z'{}^2}{F}  \label{oal}
\eeqa

\subsubsection{Periodic spike solution}

%Our aim is to consider a special  solution for which
%In
Besides
 $x_+=\tau$ let us now  make the additional gauge choice $x_-=\sigma$ and
 look for a solution
corresponding to a rigid string moving along $x=\sqrt 2 (x_+ + x_-)$ with
constant velocity $v$.
 This implies that
\beq
z=z(\xi),
\ \ \ \ \ \ \
\xi\equiv x-vt
=x_+-\frac{v+1}{v-1}x_- = \tau-\frac{1}{\eta_0^2}\sigma, \ \
\ \ \ \  \mbox{with}\ \  \eta_0^2\equiv \frac{v-1}{v+1}
\eeq
We defined $\eta_0$ taking into account that, for our solution, $v>1$.
Notice that this does not imply that the string moves faster than
light since its  actual speed should be measured using the bulk metric.
In fact, this  implies that the string does not reach the boundary.

With this ansatz, the $x_-$  equation of motion  becomes
\beq
 \partial_\xi\left(\frac{1}{\eta_0^2z^2F}\right) = 0
\eeq
 which implies
\beq
\partial_\xi z=\frac{\eta_0^2}{\mu z^2}\sqrt{\frac{z_0^4-z^4}{z^2-z_1^2}}\ , \ \ \ \ \ \ \ \ \
z_1 \equiv \sqrt{2}\,\frac{\eta_0}{\mu}
\eeq
where
$z_0$ is a constant of integration.

\begin{figure}
\epsfig{file=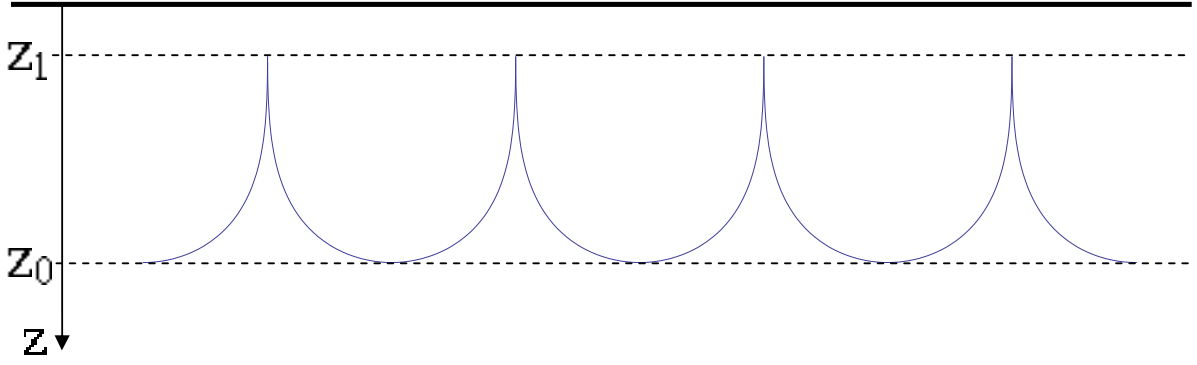, width=16cm}
\caption{Periodic spike solution for a string in  AdS--pp-wave metric.}
\label{fig1}
\end{figure}

After gluing it with its copies under  translations and reflections,
the solution takes
shape shown in fig.\ref{fig1}.
Using the equation (\ref{oal}) it is straightforward to compute the conserved quantities:
\beqa
P_- &=& \frac{2T}{\mu z_0^2} \int_{z_1}^{z_0} \frac{dz}{z^2}\sqrt{\frac{z_0^4-z^4}{z^2-z_1^2}} \\
P_+ &=& -\frac{2\mu T}{z_0^2} \int_{z_1}^{z_0} dz
                     \Big[z^2\sqrt{\frac{z^2-z_1^2}{z_0^4-z^4}}+\left(1-\frac{z_1^2}{2z^2}\right)
\sqrt{\frac{z_0^4-z^4}{z^2-z_1^2}}\Big]
\eeqa
which are given in terms of the position of the valleys $z_0$ and the spikes $z_1$. The integrals
can be explicitly done in terms of the elliptic functions:
\beqa
P_- &=& \frac{2T}{\mu z_0^2} \frac{1}{b\sqrt{1+b}}\Big[-b\eK(p)+(1+b)\eE(p)-b^2\Pi(1-b,p)\Big] \\
P_+ &=& \frac{\mu T}{\sqrt{1+b}}\Big[(1+b)\eE(p)-(2+b)\eK(p)+b^2\Pi(1-b,p)\Big]
\eeqa
where
\beq
p=\sqrt{\frac{1-b}{1+b}}, \ \ \ \ \ \ \  \ b = \frac{z_1^2}{z_0^2}
\eeq
Notice that when $b\rightarrow 0$ we get
\beqa
 P_- \simeq& \frac{2T}{\mu z_0^2}\frac{1}{b} \ , \ \ \ \ \ \ \ \ \ \ \ \
 P_+ \simeq& \mu T \ln b
\eeqa
 so that
\beq
 P_+ \simeq -\mu T \ln P_-
\eeq
 which is the expected result \ci{ppwave}  at the  leading order in the large spin
limit $S\rightarrow\infty$.

Finally, the separation $\Delta x_-$ (for constant $x_+$) between spikes can be computed as
\beqa
 \Delta x_- &=& \int d\sigma = \eta_0^2 \int d\xi = 2\mu \int_{z_0}^{z_1} z^2\, dz \sqrt{\frac{z^2-z_1^2}{z_0^4-z^4}} \\
            &=& \frac{\mu z_0^2}{\sqrt{1+b}} \Big[(1+b) \eE(p) - b \eK(p) - b^2 \Pi(1-b,p)\Big]
\eeqa
This allows us to compute
\beqa
 P_- \Delta x_- &=& \frac{2T}{b(1+b)} \Big[(1+b) \eE(p) - b \eK(p) - b^2 \Pi(1-b,p)\Big]^2 \\
 P_+ &=& \frac{\mu T}{\sqrt{1+b}}\Big[(1+b)\eE(p)-(2+b)\eK(p)+b^2\Pi(1-b,p)\Big]
\eeqa
We observe  that  if we set $P_- \Delta x_-  = \pi \bPp$, $P_+ = \mu \bar{\gamma}$
these expressions match the ones in eq.(\ref{res1}),
as  claimed.

The reason for  this $\Delta x_-$ factor can be understood as follows.
In the limit we consider $n \to \infty$ with $(E+S)/n^2$ being fixed.
The number of spikes  is related to the angle  difference between the
spikes that scales as  $\Delta \theta\sim  {1 \ov n}$.
Then $(E+S)/n^2 = (E+S)/n   \times  \Delta \theta$  translates into
$
P_-   \Delta x_-$ in the pp-wave picture    since here $P_{+}= E-S, \  P_{-}= -(E+S) $.

\section{Straight string in  AdS--pp-wave background}

The  straight folded  (``2-spike'') string rotating in $AdS_5$  is related, in the large spin limit,
to  the  following   simpler solution for a string
in  $AdS$--pp-wave background: the string moves along spatial $x= \sqrt 2 (x_+ + x_-)$ direction  and is
  extended  along $z$ from a finite distance  from a boundary to the horizon, i.e.
$\sigma_1\le z<\infty$ \   ($\sigma_1$ is a given constant
related to spin of solution in $AdS_3$). If $\sigma_1$ is non-zero the string does not touch the
boundary. The profile of this solution is presented in figure 2.
\begin{figure}
\epsfig{file=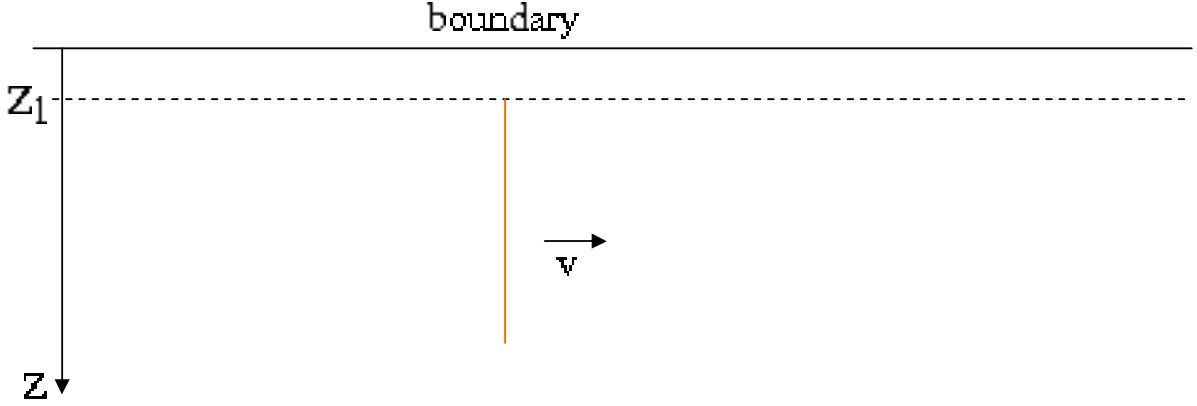, width=16cm}
\caption{Straight string solution in  AdS--pp-wave metric. The string extends all the way to infinity in $z$.}
\label{fig2}
\end{figure}

 The corresponding ansatz for string coordinates  is
\begin{equation}
x_{+}=\tau, \qquad x_{-}= V \tau, \qquad z= f(\sigma)
\end{equation}
The Nambu action is
\begin{equation}
I= - \int d \tau d \sigma\ L, \qquad \qquad L =  T \frac{f'}{f}\sqrt{\mu^2 f^2 - 2 V}
\end{equation}
The simplest solution representing a ``hanging'' string is
\be z=\sigma \ , \ \ \ \ \ \ \ \ \ \ \
\sigma_1 \leq \sigma < \infty \ , \ \ \ \ \ \ \ \
z_1 \equiv \sigma_1 =  \frac{\sqrt{2V}}{\mu} \ . \ee
One can check that
all equations of motion are satisfied in this case.

 When the ``hanging'' string touches the boundary, i.e. when
 $\sigma_1=0$  (i.e. when $V=0$ and thus the  string moves along $x_-$ only)
  it was shown in \cite{ppwave} that such a
 solution reproduces  $\frac{1}{4}$ of the leading large spin asymptotics of
  the energy of the spinning folded string  in $AdS_3$.
Here we shall consider a more general case   when   the string is not
  touching the boundary but is close to it, i.e. $\sigma_1$ is small.
 % and check what part of the folded string solution it captures.
The conserved charges are
\begin{eqnarray}
P_{+}&=&- \int_{\sigma_{1}}^{\infty} d \sigma \frac{\partial L}{\partial
(\partial_{\tau} x_{+})}\to  T \mu \int_{\sigma_{1}}^{\eR} d \sigma \frac{\sigma^2-
\frac{\sigma_1^2}{2}}{\sigma^2\sqrt{\sigma^2-\sigma_1^2}}= T \mu \Big[ \ln \frac{\eR+\sqrt{\eR^2-
\sigma_1^2}}{\sigma_1}-\frac{\sqrt{\eR^2-\sigma_1^2}}{2 \eR}\Big]\nonumber\\
&=& \frac{\mu T}{2}\Big(\ln \frac{4 \eR^2}{\sigma_1^2}-1\Big) +O(\frac{1}{\eR^4})  \label{cha1}\\
P_{-}&=& -\int_{\sigma_{1}}^{\infty} d \sigma \frac{\partial L}{\partial(\partial_{\tau} x_{-})}=-\frac{T}{\mu} \int_{\sigma_{1}}
^{\infty}\frac{d \sigma}{\sigma^2} \frac{1}{\sqrt{\sigma^2-\sigma_1^2}}=-\frac{T}{\mu \sigma_1^2} \label{cha2}
\eea
where  in $P_+$
we have introduced as in \cite{ppwave} a large cutoff $\eR$ in $\sigma$,
and expanded in large $\eR$. In the limit when $\sigma_1$ is small, so that the end of the
string is close
to the boundary, we find  that both $|P_{-}|$ and $P_{+}$ are large.
Expressing   $\sigma_1$ in terms of $P_{-}$  we obtain
\begin{equation}
P_{+}= \frac{\mu T}{2}\ln |P_{-}| + \frac{\mu T}{2}\Big( \ln \frac{4 \eR^2 \mu}{T}-1\Big)   \label{ami}
\end{equation}
As was shown in \cite{ppwave} the relation to the folded string in $AdS_3$ with energy $E$ and spin $S$
%(or spiky string with two spikes) should be done by formally identifying
can be established  by formally identifying
\begin{equation}
P_{+}= E-S, \qquad \qquad P_{-}= -(E+S)
\end{equation}
Then     using that   $S= - \frac{1}{2}(P_{-}+P_{+})$  in    (\ref{ami})
 one obtains
\begin{equation}
S= \frac{1}{2} |P_{-}|-  \frac{\mu T}{4} \ln |P_{-}| - \frac{\mu T}{4}\Big( \ln \frac{4 \mu \eR^2}{T}-1\Big)
\end{equation}
or, inverting this relation,
\begin{equation}
|P_{-}|= 2 S + \frac{\mu T}{2}\ln (2 S) + \frac{\mu T}{2}\Big( \ln \frac{4 \mu \eR^2}{T}-1\Big)
+ \frac{\mu^2 T^2}{8}\frac{\ln S}{S}+\frac{\mu^2 T^2 (\ln \frac{8 \mu \eR^2}{T}-1)}{8 S}+...  \label{wiu}
\end{equation}
Inserting $P_{-}$ back into (\ref{ami}) and expanding in large $S$ gives
\begin{eqnarray}
P_{+}&=& E-S = \frac{\mu T}{2}\ln S + \frac{\mu T}{2}\Big(\ln \frac{8 \mu \eR^2}{T}-1\Big)
+ \frac{\mu^2 T^2} {8 S}
 \Big(\ln S + \ln \frac{8 \mu \eR^2}{T}-1\Big) \label{oii}\\
&-& \frac{\mu^3 T^3 }{64 S^2}\bigg[\ln^2 S + 2 \ln S (\ln \frac{8 \mu \eR^2}{T}-2)+\ln \frac{\mu \eR^2}{T}(\ln
\frac{64 \mu \eR^2}{T}-4)+3 + 3 \ln 2 (3 \ln 2 -4)\bigg] +...
%O(\frac{\ln^3 S}{S^3})
 \nonumber
\end{eqnarray}
The leading $\ln S$ term  here is  (with the pp-wave scale  parameter set to be  $\mu=1$)
 the same as  $\frac{1}{4}$ of the $\ln S$ term  in  the folded string energy
 \cite{ppwave}.
 To compare  higher  order  terms  let us formally  replace $T  \rightarrow 4 T$ with $\mu=1$.
  We then obtain from \rf{oii}  ($T=\frac{\sqrt{\lambda}}{2 \pi}$)
\begin{eqnarray}
&& E-S = \frac{\sqrt{\lambda}}{ \pi}\ (\ln S + a)
%\frac{\sqrt{\lambda}}{\pi} a
%[\ln \frac{4 \pi \eR^2}{\sqrt{\lambda}}-1]
+ \frac{\lambda}{2 \pi^2}\frac{\ln S + a
%\ln \frac{4 \pi \eR^2}{\sqrt{\lambda}}-1
}{S}\label{aml}
- \frac{\lambda^{{3}/{2}}}{8 \pi^3 }\ {( \ln S  +a) ( \ln S  +a -2) \ov S^2}
%(\ln \frac{4 \pi \eR^2}{\sqrt{\lambda}}-2)
%\ln S +  a^2 -2 a
%\ln \frac{4 \pi \eR^2}{\sqrt{\lambda}}-
%a-3)(a+1)
%\ln \frac{4 \pi \eR^2}{\sqrt{\lambda}}
+
 O(\frac{\ln^3 S}{S^3})\nonumber  \\
&& \ \ \ \ \ \ \ \ \ \ \ a\equiv \ln \frac{4 \pi \eR^2}{\sqrt{\lambda}}-1
\end{eqnarray}
This  may be  compared to  the corresponding expression
 for the  classical energy of  the folded string in $AdS_3$  \cite{bftt}
 \begin{eqnarray}
&& E-S = \frac{\sqrt{\lambda}}{ \pi}\ (\ln S + b)
%\frac{\sqrt{\lambda}}{\pi} a
%[\ln \frac{4 \pi \eR^2}{\sqrt{\lambda}}-1]
+ \frac{\lambda}{2 \pi^2}\frac{\ln S + b
%\ln \frac{4 \pi \eR^2}{\sqrt{\lambda}}-1
}{S}\label{amlk}
- \frac{\lambda^{{3}/{2}}}{8 \pi^3 }\ {( \ln S  +b) ( \ln S  +b -{5 \ov 2}) -1 \ov S^2}
%(\ln \frac{4 \pi \eR^2}{\sqrt{\lambda}}-2)
%\ln S +  a^2 -2 a
%\ln \frac{4 \pi \eR^2}{\sqrt{\lambda}}-
%a-3)(a+1)
%\ln \frac{4 \pi \eR^2}{\sqrt{\lambda}}
+
 O(\frac{\ln^3 S}{S^3})\nonumber  \\
&& \ \ \ \ \ \ \ \ \ \ \ b\equiv \ln \frac{8 \pi }{\sqrt{\lambda}}-1
\end{eqnarray}
%   once we take  the ``$pp$-wave limit'' in the latter, i.e.
%   after rescaling $v \rightarrow \epsilon^2 v$,
%  where $v=\frac{1}{\cosh 2 \rho}$ (see section 2).
The two expressions do have the same structure.
We observe that the coefficients of the terms
$\ln S,\ \frac{\ln^k S}{S^k}$ in (\ref{aml}) that do not
 depend on the cutoff $\eR$  do match.
  However, the coefficients of some of
  the subleading  $\frac{1}{S^k}$ terms  which depend on  $\eR$
  appear to disagree. This is  not surprising  since
   we cannot  unambiguously  fix the cutoff in (\ref{ami}).

 This partial matching can be understood as a consequence of the fact that in the large $S$ limit
 the coefficients of the leading terms at each order in
  $\frac{1}{S^k}$ receive contributions  from the region of the folded string where
 $\rho$ is large, while the subleading terms are sensitive to the smaller   $\rho$ region.
 % of the folded string far form the large $\rho$ region.
  The latter is  not ``seen'' in the ``$pp$-wave limit''
  where one zooms in at the near-boundary part of the  $AdS_5$.

 %   the  comparison with the
 %   folded string in the $pp$-wave limit is not possible.
 %   We note still the similar
 %   structure of (\ref{aml}) with the full folded string solution result (\ref{qml}).
% Let us recall the large $S$ expansion of the classical energy of the folded string \cite{bftt} (full expression, not in the $pp$-wave limit)
% \begin{eqnarray}
% E-S&=&\frac{\sqrt{\lambda}}{ \pi}\ln S + \frac{\sqrt{\lambda}}{\pi}[\ln \frac{8 \pi }{\sqrt{\lambda}}-1]+\frac{\lambda}{2 \pi^2}\frac{\ln S + \ln \frac{8 \pi}{\sqrt{\lambda}}-1}{S} \label{qml}\\
% &-& \frac{\lambda^{\frac{3}{2}}}{8 \pi^3 S^2}\bigg[\ln^2 S +2(\ln \frac{8 \pi}{\sqrt{\lambda}}-\frac{9}{4}) \ln S +\frac{5}{2}+(\ln \frac{8 \pi}{\sqrt{\lambda}}-\frac{9}{2})\ln \frac{8 \pi}{\sqrt{\lambda}}\bigg]+ O(\frac{\ln^3 S}{S^3}) \nonumber
% \end{eqnarray}
%An interesting fact is that the terms $\ln S, \frac{\ln^k S}{S^k}$ of the folded string are precisely
%captured by the expression (\ref{aml}).

% As in \cite{bftt}, the $\ln^k S/S^k$ terms can be organized in terms of a simpler function, then (\ref{qml1}) can be written as
% \begin{equation}
% E-S =  \frac{\sqrt{\lambda}}{\pi} \ln [ S+ \frac{1}{2}\frac{\sqrt{\lambda}}{\pi}\ln S] + ...\label{fhj}
% \end{equation}
% where the last dots represent contributions do not contain $\ln S$ and $\frac{\ln^k S}{S^k}$ type contributions.

\section{Spiky strings in $AdS_5$  with angular momentum in $S^5$:\\ General relations}

Let us now generalize the spiky string  solution of \cite{spiky} to the case of non-zero semiclassical
 angular momentum $J$  and winding $m$  in $S^5$. This should be  important for a detailed comparison with the
 states  in the $SL(2)$ sector  on  the gauge theory side, i.e. the strings should   carry
spin $S$ in $AdS_3$ and spin $J$   in $S^1\subset S^5$.

Let us start with the $AdS_3 \times S^1$ metric in  embedding  coordinates
and, following  \ci{krt1}, use the  conformal gauge.
Then the string  Lagrangian  takes the form
%\begin{equation}
%ds^2=-dY_0 dY_{0}^{*}+dY_1 dY_{1}^{*}+dX dX^{*}
%\end{equation}
%We shall follow \ci{krt1}   and   work in conformal gauge. The Lagrangian is
\begin{equation}
L=-\frac{1}{2}\bigg[-\partial_a Y_0 \partial^a Y_0^* + \partial_a Y_1 \partial^a Y_1^*+\partial_a X \partial^a X^*
+\Lambda (| Y_0|^2-|Y_1|^2-1) + \tilde{\Lambda}(|X|^2-1)\bigg]
\end{equation}
The conformal constraints are
\begin{equation}
-|\dot{Y}_0|^2+|\dot{Y}_1|^2+|\dot{X}|^2-|Y'_0|^2+|Y'_1|^2+|X'|^2=0,
\quad -\dot{Y}_0 Y_0^{'*}+ \dot{Y}_1
Y_1^{'*}+\dot{X}X^{'*}+c.c.=0
\end{equation}
We shall consider the following rigid string
ansatz which is similar to the one used in the  $R \times S^5$  case  in \cite{krt1}
\bea
&&Y_0=y_0(u)\ e^{i w_0 \tau}\ , \qquad \quad Y_1=y_1(u)\ e^{i w_1 \tau}\ ,
\qquad \quad X=x(u)\ e^{i \nu \tau}\ ,\\
&& \quad \quad u\equiv \alpha
\sigma+\beta \tau
\eea
With this ansatz the string Lagrangian reduces to the following 1-dimensional
integrable  \ci{ft3,krt1} mechanical system
(here prime is derivative over the argument $u$)
\begin{eqnarray}
L&=&-\frac{1}{2}\bigg[(\beta^2-\alpha^2)(|y'_0|^2 -|y'_1|^2
-|x'|^2)+w_0^2 |y_{0}|^2-w_1^2 |y_1|^2-\nu^2 |x|^2
+i w_0 \beta (y_0 y_0^{'*}-y_0^{*}y'_{0})\nonumber\\
&-&i w_1 \beta (y_1 y_1^{'*}-y_1^{*}y'_{1})-i\beta \nu(x
x^{'*}-x'x^{*})+\Lambda(|y_0|^2-|y_1|^2-1)+ \tilde{\Lambda}(|x|^2-1) \bigg]\la{onee}
\end{eqnarray}
The corresponding conserved Hamiltonian is
\begin{equation}
H=\frac{1}{2}\bigg[-(\beta^2-\alpha^2)(|y'_0|^2-|y'_1|^2-|x'|^2)+w_0^2
|y_0|^2-w_1^2 |y_1|^2-\nu^2 |x|^2\bigg]
\end{equation}
After combining the two conformal constraints they can be
written as
\begin{equation}
(\beta^2-\alpha^2)(|y'_0|^2-|y'_1|^2-|x'|^2)-w_0^2 |y_0|^2+w_1^2
|y_1|^2+\nu^2 |x|^2=0  \label{c1}
\end{equation}
\begin{equation}
\frac{\beta^2-\alpha^2}{2\beta}(-w_0 \xi_{0}+w_1 \xi_1+\nu
\xi_{2})-w_0^2 |y_0|^2+w_1^2|y_1|^2+\nu^2 |x|^2=0  \label{c2}
\end{equation}
where
\begin{equation}
\xi_{0}=i(y_0 y_0^{*'}-y'_0 y_0^{*}), \quad \xi_{1}=i(y_1
y_1^{*'}-y'_1 y_1^{*}), \quad \xi_{2}=i(x x^{*'}-x'x^{*})
\end{equation}
The first constraint (\ref{c1}) is conserved since it is just equivalent to $-2 H=0$, while
the second one (\ref{c2}) is satisfied due to the equations of motion.
The equations of motion for $y_0,y_1,x$ imply
\begin{equation}
(\beta^2-\alpha^2) \xi_0' = - 2 w_0 \beta (y_0 y_0^*)', \quad  (\beta^2-\alpha^2) \xi_1' = - 2 w_1 \beta (y_1 y_1^*)', \quad
(\beta^2-\alpha^2) \xi_2' = - 2 \nu \beta (x x^*)'\nonumber
\end{equation}
Since we consider a closed  string,  the condition of periodicity in $\sigma$ implies periodicity in $u$
\begin{equation}
y_0(u)=y_0(u+ 2 \pi \alpha), \quad y_1(u)=y_1(u+ 2 \pi \alpha), \quad x(u)=x(u+ 2 \pi \alpha)  \label{per}
\end{equation}
Since  $y_0,y_1$ are in general complex  and $|y_0|^2 - |y_1|^2=1, \ |x|^2=1$   we may set
%  now further specify our ansatz as follows
%We consider further the ansatz
\begin{equation}
y_0=r_0(u)e^{i \varphi_0(u)}, \qquad \quad y_1=r_1(u)e^{i
\varphi_1(u)},  \qquad \quad x=e^{i \psi(u)} \ , \ \ \   \ \ \ \   r_0^2- r^2_1=1  \ .
\end{equation}
Then the  Lagrangian \rf{onee} becomes
\begin{eqnarray}
L&=&-\frac{1}{2}\bigg[(\beta^2-\alpha^2)(r_0^{'2}-r_1^{'2})+r_0^2(\beta^2-\alpha^2)\bigg(\varphi'_{0}+\frac{\beta
w_0}{\beta^2-\alpha^2}\bigg)^2-\frac{\alpha^2 r_0^2
w_0^2}{\beta^2-\alpha^2}\nonumber\\
&-&r_1^2(\beta^2-\alpha^2)\bigg(\varphi'_{1}+\frac{\beta
w_1}{\beta^2-\alpha^2}\bigg)^2+\frac{\alpha^2 r_1^2
w_1^2}{\beta^2-\alpha^2}-(\beta^2-\alpha^2)\bigg(\psi'+\frac{\beta
\nu}{\beta^2-\alpha^2}\bigg)^2\nonumber\\
&+&\frac{\alpha^2
\nu^2}{\beta^2-\alpha^2}\bigg]+\Lambda(r_0^2-r_1^2-1)
\end{eqnarray}
The periodicity conditions (\ref{per}) imply
\begin{equation}
r_0(u)=r_0(u+2 \pi \alpha),\ \ \ \ \ \ \  \qquad r_1(u)=r_1(u+2 \pi \alpha)
\end{equation}
\begin{equation}
 \varphi_0(u)=\varphi_0 (u+ 2 \pi \alpha) - 2 \pi m_0, \quad
\varphi_1(u)=\varphi_1(u+ 2 \pi \alpha) - 2 \pi m_1, \quad \psi(u)=\psi(u + 2\pi \alpha) - 2 \pi m\nonumber
\end{equation}
where $m_0,m_1,m$ are integers. Below we shall assume that  $m_0=0$
since we consider the global $AdS_5$ time $t$ as  non-compact.\foot{We recall that in terms of the coordinates used in \rf{mam} \  \
$Y_0= \cosh \rho \ e^{it}, \ \ Y_1= \sinh \rho \ e^{i\theta},\ \  X= e^{i \phi}$.}

%In this paper we look for solutions with no winding in $t$ that means $m_0=0$.

The equations of motions for $\varphi_{0},\varphi_{1},\psi$ can be
integrated as
\begin{equation}
\varphi'_{0}=-\frac{1}{\beta^2-\alpha^2}\bigg(\frac{C_0}{r_0^2}+w_0
\beta\bigg),  \quad
\varphi'_{1}=\frac{1}{\beta^2-\alpha^2}\bigg(\frac{C_1}{r_1^2}-w_1
\beta\bigg), \quad \psi'=\frac{D-\beta \nu}{\beta^2-\alpha^2}
\end{equation}
where $C_0,C_1,D$ are constants. The equation for $\psi$ can be
integrated again so that in $S^5$ we just have a rotating string
wound on a circle.
Denoting the angle in $S^1\subset  S^5$ by $\phi$,
%i.e.  $X=e^{i \phi}$,
 we have
\begin{equation} X=e^{i \phi}\ , \ \ \ \ \ \ \ \ \ \ \
\phi= \nu  \tau  + \psi = \nu  \tau  + \frac{(D- \beta \nu)}{\beta^2 - \alpha^2}  u
 \label{wind}
\end{equation}
The winding number in $\phi$ is defined as
\begin{equation}
2 \pi m= \int^{2 \pi \alpha}_0 du\   \psi' = \frac{D- \beta \nu}{\beta^2-\alpha^2}\int d u  \label{wal}
\end{equation}
The condition of having no winding in the $t$ direction gives the condition
\begin{equation}
2\pi m_0=\int^{2 \pi \alpha}_0 du\ \varphi_0'  = \int^{2 \pi \alpha}_0 d u \big(\frac{C_0}{1+r_1^2}+w_0 \beta \big)=0
 \label{twrap}
\end{equation}
The effective Lagrangian for $r_0,r_1$  that reproduces the remaining equations of motion is then
\begin{eqnarray}
L&=&-\frac{1}{2}\bigg[r_0^{'2}(\beta^2-\alpha^2)-\frac{C_0^2}{r_0^2(\beta^2-\alpha^2)}-
\frac{r_0^2\alpha^2w_0^2}{\beta^2-\alpha^2}-r_1^{'2}(\beta^2-\alpha^2)+\frac{C_1^2}{r_1^2(\beta^2-\alpha^2)}\nonumber\\
&+& \frac{r_1^2 \alpha^2
w_1^2}{\beta^2-\alpha^2}+\frac{D^2}{\beta^2-\alpha^2}+\frac{\alpha^2\nu^2}{\beta^2-\alpha^2}\bigg]+\Lambda(r_0^2-r_1^2-1)
\end{eqnarray}
The Hamiltonian is
\begin{eqnarray}
H&=&-\frac{1}{2}r_0^{'2}(\beta^2-\alpha^2)-\frac{C_0^2}{2r_0^2(\beta^2-\alpha^2)}-
\frac{r_0^2\alpha^2w_0^2}{2(\beta^2-\alpha^2)}+\frac{1}{2}r_1^{'2}(\beta^2-\alpha^2)+\frac{C_1^2}{2r_1^2(\beta^2-\alpha^2)}\nonumber\\
&+& \frac{r_1^2 \alpha^2
w_1^2}{2(\beta^2-\alpha^2)}+\frac{D^2}{2(\beta^2-\alpha^2)}+\frac{\alpha^2\nu^2}{2(\beta^2-\alpha^2)}
\end{eqnarray}
The first constraint (\ref{c1}) is just
\begin{equation}
-2H =0  \label{con1}
\end{equation}
The second constraint (\ref{c2}) gives the condition
\begin{equation}
w_0 C_0+w_1 C_1+D\nu=0  \label{con2}
\end{equation}
In the special case of  $D=0$ to which we shall return  below
 the constants $C_0$ and $C_1$ will have opposite signs as we
will assume without loss of generality $w_0,w_1$ are positive
 (one  may assume that  $C_1$ is  negative).

The conserved charges are
% ($T=\frac{\sqrt{\lambda}}{2 \pi}$)
\begin{equation}
E= T\int \frac{d u}{\alpha} \bigg(-\frac{\beta C_0}{\beta^2-\alpha^2}-\frac{\alpha^2 \nu_0 (1+ r_1^2)}{\beta^2-\alpha^2}\bigg)
\end{equation}
\begin{equation}
S= T \int \frac{d u}{\alpha}\bigg(\frac{\beta C_1}{\beta^2-\alpha^2}-\frac{\alpha^2 w_1 r_1^2}{\beta^2-\alpha^2}\bigg), \quad
J= T \int \frac{d u}{\alpha}\bigg(\frac{D \beta}{\beta^2-\alpha^2}-\frac{\alpha^2 \nu}{\beta^2-\alpha^2}\bigg)
\end{equation}
 % in this particular case.
Using that $r_0^2-r_1^2=1$ we get from (\ref{con1}) the equation for $r_1$
\begin{equation}
(\beta^2-\alpha^2)^2
r_1^{'2}=(1+r_1^2)\bigg(\frac{C_0^2}{1+r_1^2}+\alpha^2 w_0^2
(1+r_1^2)-\frac{C_1^2}{r_1^2}-\alpha^2 w_1^2 r_1^2 -D^2-\alpha^2
\nu^2\bigg) \label{r1}
\end{equation}
Let us  first look for  solutions with two turning points ($r_1'=0$)
 at some finite values of $r_1$  (the idea is that this should  represent an ark of the spiky string).
Considering (\ref{r1}) at large $r_1$, we observe that we need to
satisfy the condition $w_0^2 < w_1^2$ in order for the string  not to reach the boundary.
% i.e. we are interested here in a solution within a finite range in $\rho$.

Let us express the  equation (\ref{r1}) in terms of the variable $v$
\begin{equation}
v= \frac{1}{1+ 2 r_1^2}=\frac{1}{\cosh 2 \rho}, \qquad \qquad 0 \leq v \leq 1
\end{equation}
where we used that in  terms of global $AdS_3$ coordinate $\rho$ we have
 $r_1= \sinh \rho$.
 %, so that  $v=\frac{1}{\cosh 2 \rho}$.
%Equation (\ref{r1}) then becomes
Then  \begin{eqnarray}
(\beta^2-\alpha^2)^2 v'^2&=& 2 v [4 C_0^2 v^2 (1-v)+ \alpha^2 w_0^2 (1-v) (1+v)^2-4 C_1^2 v^2 (1+v) \nonumber\\
&-& \alpha^2 w_1^2 (1-v)^2 (1+v) -2 (D^2+\alpha^2 \nu^2) v (1-v^2)]  \label{v1}
\end{eqnarray}
Without loss of generality one can set  $\alpha$ in $u$ to any given value
 ($\alpha$ can be absorbed into other parameters).
In what follows we shall
assume   $\alpha=1$.
Then the equation (\ref{v1}) becomes
\begin{equation}\la{vew}
v'= \frac{\sqrt{2 v P(v)}}{1-\beta^2}
\end{equation}
where
\begin{eqnarray}
P(v)&=&v^3 [-4 C_0^2 -4 C_1^2 +2 (D^2+ \nu^2) -w_0^2-w_1^2]+v^2 (4 C_0^2-4 C_1^2-w_0^2+w_1^2)\nonumber\\
&+&v [w_0^2+w_1^2-2 (D^2+\nu^2)] +w_0^2-w_1^2  \label{poli}\\
&\equiv &
 [-4 C_0^2 -4 C_1^2 +2 (D^2+ \nu^2) -w_0^2-w_1^2]\  (v-v_1) (v-v_2) (v-v_3)
\eea
where $v_n$ are three roots of $P(v)=0$.
To have a  consistent string solution all roots should be real.
%In order to have string solutions we need two real positive roots; this means we want to have all three solutions of the equation
%$P(v)=0$ real.
We should also take into account the  conditions (\ref{con2}),(\ref{wal}),(\ref{twrap}) which
may  be used to eliminate some of the constants in terms of the other constants.

Let us assume that $P(v)$ has two positive roots $0 \leq v_2 \leq v_3 \leq 1 $, and one negative $v_1 \leq 0$
 (as we shall  see below the  constants $C_0, C_1$
 can be always  chosen so that this is true).
 The product of the roots is determined by the parameter
\begin{equation}
a \equiv -4 C_0^2 - 4 C_1^2 +2 (D^2+ \nu^2) - w_0^2- w_1^2=\frac{w_1^2 -w_0^2}{v_1 v_2 v_3}  \label{qho}
\end{equation}
We note that then $a <0 $ which means that between the two positive roots $v_2, v_3$ the polynomial $P(v)$
is positive. This is our range of interest,
meaning that  for a physical solution with two turning points  we have
 $v_2 \leq v \leq v_3$. The two physical constants that we
 are to fix are $v_2, v_3$. One can then find $v_1, C_0, C_1$ in terms of $v_2, v_3$.
 %We can write the polynomial as
 We can write the polynomial as
\begin{equation}
P(v)= \frac{w_1^2 -w_0^2}{v_1 v_2 v_3}(v-v_1)(v-v_2)(v-v_3)=-8 C_1^2 \frac{(v-v_1)(v-v_2)(v-v_3)}{(1-v_1)(1-v_2)(1-v_3)}
\end{equation}
where now it is understood that $v_1$ is not arbitrary but  is
 a function of $v_2,v_3$:
 \begin{equation}
v_1= - \frac{v_2 v_3}{v_2+ v_3 + v_2 v_3 \frac{w_0^2+w_1^2 -2 (\nu^2+D^2)}{w_0^2-w_1^2}}  \label{v11}
\end{equation}
 %An equivalent way to write the polynomial is
%\begin{equation}
%P(v)=-8 C_1^2 \frac{(v-v_1)(v-v_2)(v-v_3)}{(1-v_1)(1-v_2)(1-v_3)}
%\end{equation}
The expressions for the constants $C_0,C_1$ in term of $v_1,v_2,v_3$ are
\begin{equation}
C_0^2= \frac{w_0^2-w_1^2}{8}\frac{(1+v_1)(1+v_2) (1+v_3)}{v_1 v_2 v_3}, \quad
C_1^2= \frac{w_0^2-w_1^2}{8}\frac{(1-v_1)(1-v_2)(1-v_3)}{v_1 v_2 v_3}  \label{cis}
\end{equation}
%while $v_1$ can be expressed in terms of $v_2,v_3$ and other constants as
We observe that for a solution satisfying
 $-1 \leq v_1 \leq 0 \leq v_2 \leq v_3 \leq 1 $,
 we have $C_0^2 \geq 0$ and $C_1^2 \geq 0$, i.e.  our
 choice of roots  of $P(v)$ is indeed consistent.

To get  solutions with  $n$ spikes we need to
 glue together a number of $2 n$ pieces of integrals between a
 minimum ($v_2$) and a maximum  ($v_3$). In other words, wherever it appears,
 the integral $\int d u$ is to be replaced by
\begin{equation}
\int d u = 2n \int_{v_2}^{v_3} \frac{d v }{v'}= \frac{2 n (1-\beta^2)}{\sqrt{-2 a}}I_1
\end{equation}
where $I_1$ is defined in Appendix.
The winding number $m$ in (\ref{wal}) becomes
\begin{equation}
m = \frac{\beta \nu -D}{\pi \sqrt{-2 a}} n I_1  \label{sir}
\end{equation}
Solving for $D$ and using (\ref{con2}) we obtain the equation for $\nu$
\begin{equation}
w_0 C_0 + w_1 C_1 + \nu (\beta \nu - \frac{\pi m \sqrt{-2 a}}{ n I_1} )=0
\end{equation}
The condition (\ref{twrap}) gives an  additional relation between the constants, which allows
to eliminate one of them, for example,   $\beta$
\begin{equation}
2 C_0 I_5 +  w_0 \beta I_1=0  \label{cak}
\end{equation}
where $I_5$ is defined in Appendix.

For solutions with $n$ spikes the conserved charges are
\begin{equation}
\frac{\pi \mathcal{E}}{ n}=\frac{\beta C_0}{\sqrt{-2 a}}I_1 + \frac{w_0}{2 \sqrt{-2 a}}I_3, \quad \frac{\pi \mathcal{S}}{
n}=-\frac{\beta C_1}{\sqrt{-2 a}}I_1 + \frac{w_1}{2 \sqrt{-2 a}}I_2, \quad \frac{\pi \mathcal{J}}{ n}=\frac{\nu - \beta D}{\sqrt{-2 a}}I_1  \label{aiu}
\end{equation}
where the integrals are defined in Appendix and can be written in terms of the
elliptic integrals. Here  $\E = 2\pi T \mathcal{E}, \quad  S= 2\pi T \mathcal{S},  \quad J= 2\pi T \mathcal{J}$.
%\begin{equation}
%E= T \mathcal{E}, \quad \quad  S= T \mathcal{S}, \quad \quad J= T \mathcal{J}
%\end{equation}

The cartesian coordinate $Y_1$ can be expressed as
\be Y_1= \sinh \rho\ e^{i \theta} \ , \ \ \ \ \ \ \ \ \ \ \ \ \ \ \ \     \theta= w_1 \tau +
\int du\ \varphi_1' \ee
The number of spikes can be introduced  via  $\Delta \theta = \frac{2 \pi}{2 n}$ at fixed
\be t= w_0 \tau + \varphi_0(u) \ee
Here $\Delta
 \theta$ is the angle between a minimum (valley) and a maximum ($I_6$ is again defined in Appendix)
\begin{equation}
\Delta \theta = \int d \theta= \frac{1}{\beta^2-\alpha^2}\int d u \big(\frac{C_1}{r_1^2}+ \frac{w_1}{w_0}
\frac{C_0}{1+r_1^2}\big)= -\frac{2}{\sqrt{-2 a}}(C_1 I_6 + \frac{w_1}{w_0}C_0 I_5)  \label{nsp}
\end{equation}
%where  $I_6$ is defined in appendix A.
To see whether the spikes end in cusps or not we need  to evaluate the
 derivative  at the maximum value in $\rho$ or minimum  value of $v=v_2$ with $t$ fixed
\begin{equation}
\frac{d \rho}{d \theta}\bigg|_{v=v_2} = \frac{\rho' d u}{w_1 d \tau + \varphi_1' d u}\bigg|_{v=v_2}
\end{equation}
Using that for fixed $t$ we have  $d t=w_0 d \tau + \varphi_0' d u=0$ we get
\begin{equation}
\frac{d \rho}{d \theta}\bigg|_{v=v_2} = \frac{\rho'}{\varphi_1' - \frac{w_1}{w_0}\varphi_0'}\bigg|_{v=v_2}  \label{qko}
\end{equation}
Evaluating this expression we obtain
\begin{equation}
\frac{d \rho}{d \theta}\bigg|_{v=v_2} =  \frac{\sqrt{P(v)} \sqrt{1-v^2}}{\sqrt{2}v^{\frac{3}{2}}} \frac{w_0 w_1}{w_0^2+w_1^2}\frac{1}{\frac{w_1^2-w_0^2}{w_0^2+w_1^2}-v}\bigg|_{v=v_2}  \label{spikes}
\end{equation}
We observe that while $P(v_2)=0$, the denominator does not vanishes in general. For the particular case when $v_2= \frac{w_1^2-w_0^2}{w_0^2+w_1^2}$ the denominator does vanish but this case corresponds to the situation when the motion is only in $AdS_5$ (see below).
Thus in general when
the string is moving or extended in
 $S^5$   the spikes at $v=v_2$ are rounded, i.e. they do not end in cusps.

To illustrate   the rounding of  spikes we set for
simplicity $D=0$ while keeping $m, \nu$ non-zero, and compute the
 angle $\theta$ in terms of $v_1,v_2,v_3$ at fixed $t$
\begin{eqnarray}
&&  \theta(v)= \frac{1}{2 \sqrt{(1-v_1)(1-v_2)(1-v_3)}}\Big[(1+v_1)(1+v_2)(1+v_3)
I_5(v_2,v)\nonumber\\
&&\ \ \ \ \ \ \ \ \ \   - (1-v_2)(1-v_2)(1-v_3)I_6(v_2,v)\Big]
\end{eqnarray}
%where $I_5(v_2,v),I_6(v_2,v)$ are defined in Appendix A.
For $D=0$ we have $w_0 C_0+w_1 C_1=0$. The relevant equations simplify as
\begin{equation}
\frac{w_1^2}{w_0^2}=\frac{(1+v_1)(1+v_2)(1+v_3)}{(1-v_1) (1-v_2)(1-v_3)}  \label{cond1}
\end{equation}
Using (\ref{cond1}) in (\ref{v11}) we get
\begin{equation}
\frac{\nu}{w_0}=\sqrt{\frac{(v_1+v_2)(v_1+v_3)(v_2+v_3)}{v_1 v_2 v_3 (v_1-1)(v_2-1)(v_3-1)}}  \label{cond2}
\end{equation}
Solving for  $\beta$ in  (\ref{cak}) and using this  in (\ref{sir})  gives\footnote{Note that the overall
 sign in this equation is not relevant since $m$ should be   an integer number.}
\begin{equation}
m = - \frac{\nu}{2\pi w_0} n I_5 \sqrt{(1+v_1)(1+v_2)(1+v_3)}
\end{equation}
The condition $w_1^2 > w_0^2$ along with (\ref{cond1}) implies
\begin{equation}
v_1+v_2+v_3 + v_1 v_2 v_3 \geq 0, \quad \quad v_1 \geq - \frac{v_2 + v_3}{1+ v_2 v_3}  \label{cond3}
\end{equation}
The requirement (\ref{cond3}) as well as the condition coming from the positivity of the square root
in (\ref{cond2}) imply certain
ranges for the parameters $v_1,v_2,v_3$.

There are two possible regimes.
 The first one is represented by $\frac{v_2+v_3}{1+v_2 v_3} \geq |v_1| \geq v_3$.
 This corresponds to spikes at maximal values of $\rho$. In contrast with the case  of  ``true''
 spikes when the string  motion is in $AdS_5$ only \cite{spiky} here  for nonzero $J$
 the spikes are rounded. A typical plot in this sector is shown in figure 3.
\begin{figure}
\epsfig{file=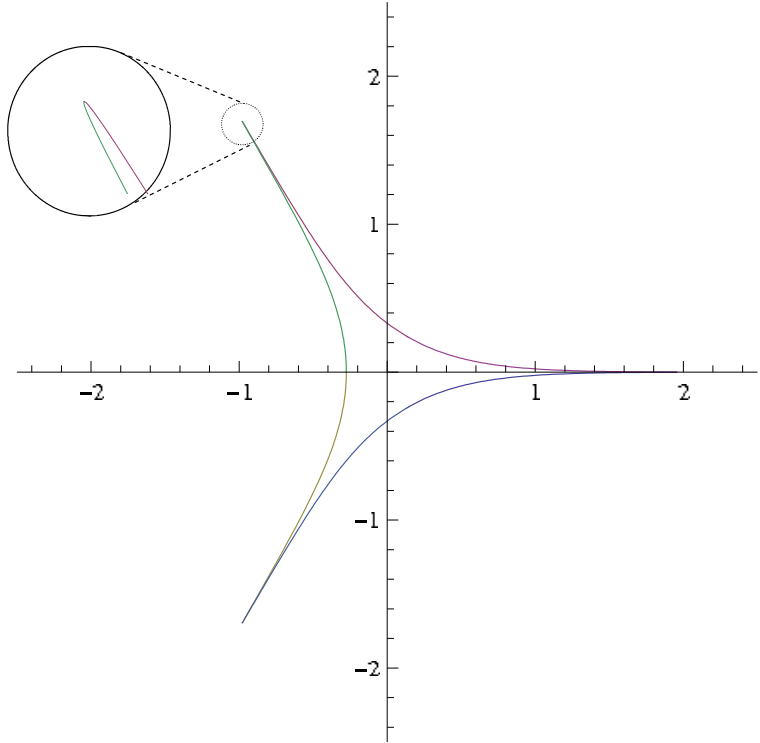, width=10cm}
\caption{Solution in polar coordinates ($\rho,\theta)$ for $n=3$ spikes for $v_1=-0.8698,
 v_2=0.04, v_3=0.865$. The   shape of the string near  maximal value of the radial coordinate
 is actually  rounded.}
\label{threespikes}
\end{figure}
The other regime is with $|v_1| \leq v_2$ and corresponds to spikes at the minimum values of $\rho$, i.e spikes in the interior. For the string moving only in $AdS_5$ this solution was found in \cite{ms}. Here again the spikes are rounded due to the presence of $J$. A typical plot is presented in figure 4.
\begin{figure}
\epsfig{file=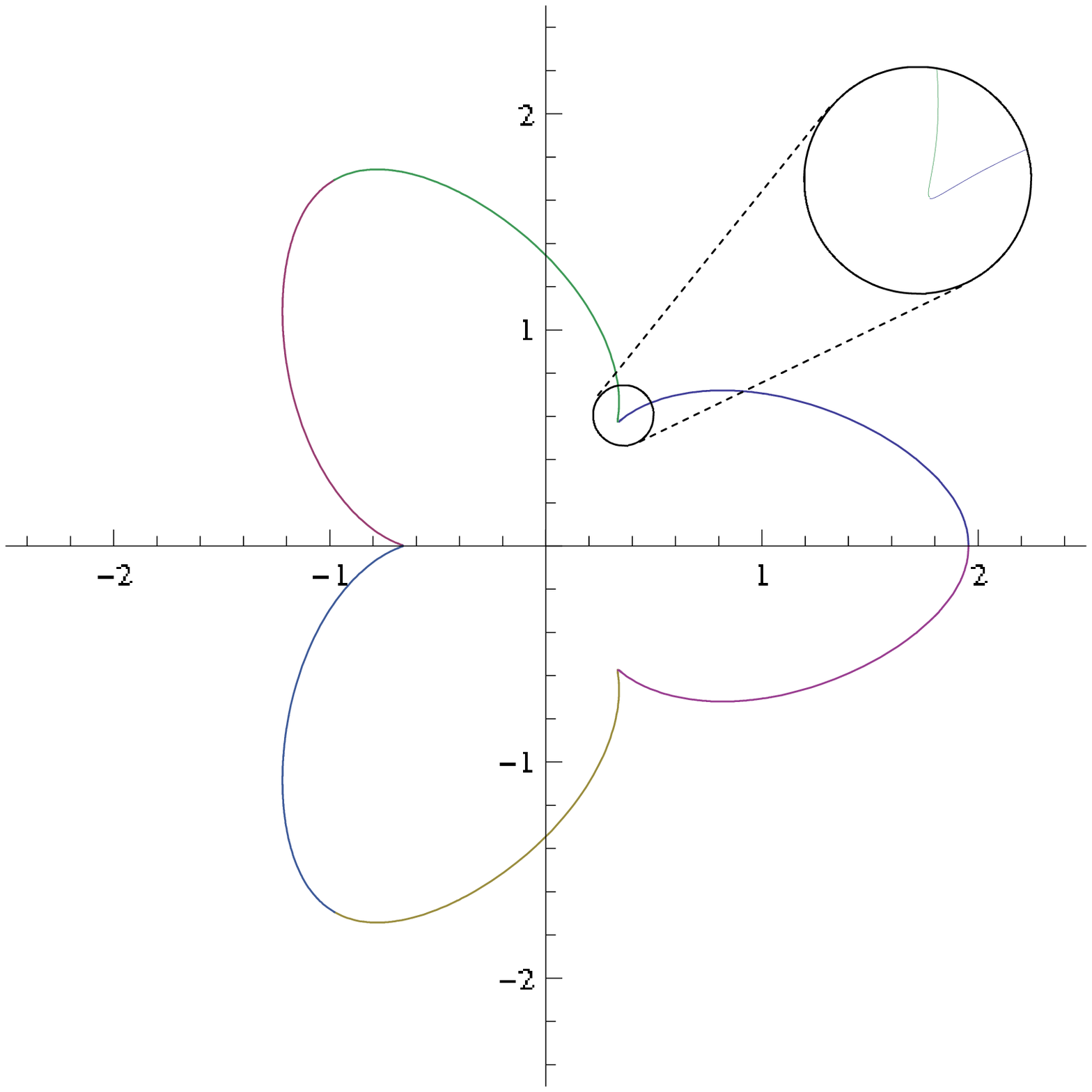, width=10cm}
\caption{Solution in polar coordinates ($\rho,\theta)$ for $n=3$ spikes for $v_1=-0.02, v_2=0.04,
 v_3=0.4957$.   The   shape of the string near  minimal  value of the radial coordinate
 is  actually  rounded.}
\label{threespikes_in}
\end{figure}

Returning to the general case with $D \neq 0$ we
 observe that the  solution is parametrized  by  only four independent parameters. For example,
  all the charges $E,S,J,m$ can be expressed in term of
  the parameters $\frac{w_1}{w_0},\frac{\nu}{w_0},\frac{D}{w_0},n,v_1,v_2,v_3$.
   Furthermore, the ratios $\frac{w_1}{w_0},\frac{\nu}{w_0},\frac{D}{w_0}$ can be
   written   in terms of   $n, v_1,v_2,v_3$ using  (\ref{qho}),(\ref{con2}),(\ref{nsp});
   alternatively, we may use the set of $n, \frac{w_1}{w_0},v_2,v_3$.
   The explicit expressions are not very illuminating  and we  will not present them here.
Given that there are  four independent parameters the energy can be written as
\begin{equation}
\mathcal{E}= \mathcal{E}(\S, \J, n, m) \ .
\end{equation}
Without making any further assumptions this expression is rather complicated
 due to the complicated way the roots $v_1,v_2,v_3$ enter the equations:
 as in  other similar cases \ci{ft3,krt1} one is  to solve a system  of
 parametric equations containing elliptic integrals.

\section{Spiky strings in $AdS_5$  with angular momentum in $S^5$:\\  \ \ \ \ \ \ \   special cases}
 %it is not possible to solve analytically for parameters explicitly and get an
  %expression for the energy $ \mathcal{E}(\S, \J, n, m)$.
  Below
  we shall look at few particular cases of the general solution of the previous section.
First, we shall demonstrate  how to reproduce the original spiky string  solution  \ci{spiky}
of section 2  in the limit when the string is not moving or stretched in $S^5$.
 % consider
% the particular limit when the entire motion is in $AdS_5$.
Another special case  is the fast string limit of large $\J$  with  $\frac{\S}{\J}$ fixed
in which one recovers  the familiar BMN-type \ci{ft2}  scaling  of the spinning string  energy.

Finally,  we  shall  consider the limit   in which   the spikes move  close
 to the boundary of $AdS_5$. This case  corresponds to a generalization of the
  ``$pp$-wave limit''  discussed in section 2 for the spiky string in $AdS_3$.

\subsection{No motion or stretching  in $S^5$}

This case   corresponds to $m=\nu=0$.
 The winding condition (\ref{wal}) then implies $D=0$. The constraint (\ref{con2}) gives
$
w_0 C_0+ w_1 C_1=0
$
which can be solved as
\begin{equation}
C_0= w_1 f, \quad \quad C_1=- w_0 f
\end{equation}
The polynomial $P(v)$  becomes
\begin{equation}
P(v)= -v^3 (1+4 f^2) (w_0^2+w_1^2) +v^2 (w_1^2-w_0^2) (1+4 f^2) + v (w_0^2+w_1^2) +w_0^2-w_1^2
\end{equation}
with the relevant choice  for the  roots being ($v_2 \leq v_3$)
%\begin{equation}
% \pm \frac{1}{\sqrt{1+4 f^2}}, \quad \quad \frac{w_1^2-w_0^2}{w_0^2+w_1^2}
%\end{equation}
%We observe that indeed one of the solutions is negative while two are positive.
% We have two choices for $v_2,v_3$ with $v_2 \leq v_3$. It turns out that the choice we want is
\begin{equation}
v_1= - \frac{1}{\sqrt{1+4 f^2}}, \quad \quad v_2 = \frac{w_1^2-w_0^2}{w_0^2+w_1^2}, \quad \quad v_3 = \frac{1}{\sqrt{1+4 f^2}}  \label{alo}
\end{equation}
Interchanging $v_2,v_3$ corresponds to another
branch for which the spikes are  at the minimal value  of  $\rho$.
In the following we focus on the the branch (\ref{alo}) corresponding  to the
spikes at the maximal value of $\rho$.

From  the condition (\ref{spikes})
we observe that in this case we  indeed have
spikes at $v=v_2$  since $\frac{d \rho}{d \theta}\big|_{v=v_2}$ blows up, as
% This is of course
 expected for
 the spiky string moving only in $AdS_5$ \cite{spiky}.

 Using the condition (\ref{cak}) to eliminate the parameter $\beta$
\begin{equation}
\beta= - \frac{2 w_1 f}{w_0} \frac{I_5}{I_1}
\end{equation}
the conserved  charges can be written as
\begin{equation}
\mathcal{E}= {n\ov 2 \pi} \frac{- 4 w_1^2 f^2 I_5 + w_0^2 I_3}{w_0 \sqrt{2 (1+4 f^2) (w_0^2 +w_1^2)}}\ , \quad \quad
\mathcal{S}={n\ov 2 \pi}
 \frac{-4 w_1 f^2 I_5 +w_1 I_2}{\sqrt{2 (1+4 f^2)(w_0^2+w_1^2)}}
\end{equation}
\begin{equation}
\frac{\pi}{n}= \frac{\sqrt{2} f(w_1^2  I_5 - w_0^2  I_6)}{w_0 \sqrt{ (1+4 f^2) (w_0^2 +w_1^2)}}
\end{equation}
In the arguments  above integrals $I_r$ (see Appendix)
one should use $v_k$ given by  the solution (\ref{alo}).

We end up with two independent parameters $\frac{w_0}{w_1},f$.
This is precisely what one needs to express
 the energy as $\mathcal{E}=\mathcal{E}(\mathcal{S},n)$.
 While  solving
 for $\frac{w_0}{w_1}$ and $f$ in terms of $\mathcal{S}$ and $n$ analytically
  is not possible in general, one can  solve the
   parametric  equations perturbatively, e.g.,
    for large $\mathcal{S}$.

The large $\mathcal{S}$ limit corresponds to the small $v_2$ limit. This means $w_1 \rightarrow w_0 $. In this
 limit the integrals $I_5,I_6$ are finite.
 The number of spikes in this limit can be written as
\begin{equation}
\frac{\pi}{n}= \frac{f}{\sqrt{1+4 f^2}}(I_5-I_6)
\end{equation}
 The charges scale as
\begin{equation}
\mathcal{E}-\mathcal{S} \simeq - {n\ov 2 \pi} \ln v_2, \quad \quad \mathcal{S} \simeq \frac{n}{2\pi v_2},
\quad \quad  \mathcal{E}-\mathcal{S} \simeq {n\ov 2 \pi} \ln {\frac{\mathcal{S}}{n}}
\end{equation}
i.e.  in this limit we obtain the expected result
\begin{equation}
E-S = \frac{\sqrt{\lambda}}{2 \pi}n \ln \Big( \frac{2 \pi}{\sqrt{\lambda}} \frac{S}{n}\Big) + ...
\end{equation}
with  $n=2$ corresponding to the folded string case.

\subsection{Fast string limit}

Next, let us  consider the fast string limit, i.e. the limit of large $\J$ with $\frac{\S}{\J}$ fixed.
In this limit the string energy
is expected to match the energy obtained from  the corresponding solution in
 the Landau-Lifshitz   model \cite{ptt} (see next section).
This limit corresponds to taking  $w_0,w_1, \nu$ being large while
 the other parameters staying  finite,
 with, e.g.,  the parameters $v_1,v_2,v_3$ being
  arbitrary. A particular situation for which we found the direct analog
   in the LL model is for $D=0$.
    In this case  the constraint (\ref{con2}) gives $w_0 C_0+w_1 C_1=0$, i.e. at  the leading order
    $C_0 = -C_1$.
    Plugging this into (\ref{cis}) we obtain the relation  between the $v_1,v_2,v_3$
\begin{equation}\la{vv}
v_1= - \frac{v_2 + v_3}{1+ v_2 v_3} \ .
\end{equation}
%which is precisely the condition found in the next section for the solution in the LL model.
For the spin $S$ and winding number $m$ we find
\begin{equation}\la{nn}
S= J\ \frac{I_2}{2 I_1}, \quad \quad m = \frac{n}{2 \pi} \sqrt{1+ v_1 v_2 + v_1 v_3+ v_2 v_3}\ I_5
\end{equation}
To compute  $E-S-J$  we need subleading corrections in $w_0,w_1, \nu$  so we set
\begin{equation}
w_0=w + \delta w_0, \quad \quad w_1 = w + \delta w_1, \quad \quad \nu =w, \ \ \ \ \ \ w \gg 1
\end{equation}
%where $w$ is large and $\delta w_i$ are small corrections.
Using (\ref{qho}) we find for $\delta w_0, \delta w_1$
%the following relationships between the small corrections
\begin{equation}
2 w (\delta w_0 + \delta w_1) = \frac{w_1^2 -w_0^2}{v_1 v_2 v_3}(v_1 v_2 +v_1 v_3 +v_2 v_3), \quad \quad
2 w (\delta w_0 - \delta w_1)=w_0^2 -w_1^2  \label{wjo}
\end{equation}
Therefore in the fast string limit the parameters  $a, C_0$ and all integrals finite.\foot{In the limit
 under consideration
only the leading term in the expansion of $C_1+C_0$ is relevant. However, $C_0+C_1=0$ for $D=0$.}
Then the  energy for the  $n$-spike  solution  can be written as
\begin{equation}
\frac{\mathcal{E}-\mathcal{S}-\J}{4 \pi  n}\simeq  \frac{ \beta C_0 + \beta C_1 }{\sqrt{-2 a}}I_1+ \frac{w_0 I_3 -w_1 I_2 -2 \nu I_1}{2 \sqrt{-2 a}}
\end{equation}
%Using (\ref{wjo}) we obtain the energy
i.e. as
\begin{equation}
E-S-J  \simeq   \frac{n^2\lambda }{8 \pi^2 J } \ I_1
\Big[v_1 v_2 v_3 I_{+}-(v_1 v_2 + v_1 v_3 + v_2 v_3) I_1\Big] \label{qnp}
\end{equation}
where  $v_k$ are related to $S/J$ and $m$ via \rf{vv},\rf{nn}.
This  generalizes  a similar expression for the $(S,J)$ folded string \ci{ft1}.
Eq. (\ref{qnp}) will be reproduced from the  Landau-Lifshitz  model in the next section.

%\subsection{General case $\nu \neq 0$ and $m \neq 0$}

\subsection{$pp$-wave limit}
\label{sec}

To obtain the solution in the ``$pp$-wave limit''  $n \to \infty$
 we are  to consider the  scaling  ($k=1,2,3$)
\begin{equation}
v_k \rightarrow \epsilon^2 v_k, \ \ \ \ \ \ \ \ \   \epsilon \to 0  \label{resc}
\end{equation}
It turns out that in this  scaling limit
  $\nu$ and $D$ need to be kept fixed, while the constants $C_0,C_1$ scale as
\begin{equation}
C_0= c_0 \bigg(\frac{1}{\epsilon^2}+ \frac{v_1+v_2+v_3}{2}+...\bigg),
\quad \quad C_1= - c_0 \bigg(\frac{1}{\epsilon^2}- \frac{v_1+v_2+v_3}{2}+...\bigg)
\end{equation}
where $c_0$ is finite, and $v_k$ here are now  finite. We also want $w_0,w_1$ to scale as
\begin{equation}
\frac{w_0-w_1}{w_0}= r \epsilon^2
\end{equation}
where $r$ is finite. Using that $w_0^2-w_1^2= 8 \epsilon^2 c_0^2 v_1 v_2 v_3$ we find that
 $r= \frac{4 c_0^2}{w_0^2}v_1 v_2 v_3$. Other useful relations  are
\begin{equation}
2 (D^2 +\nu^2)-w_0^2 -w_1^2 = 8 c_0^2 (v_1 v_2 + v_1 v_3 + v_2 v_3), \quad \quad a= - \frac{8 c_0^2}{\epsilon^4}
\end{equation}
The constraint (\ref{con2}) becomes, at  the leading order in $\epsilon$,
\begin{equation}
c_0 r + c_0 (v_1+v_2 +v_3) + \frac{\nu}{w_0}D=0
\end{equation}
and  can be used to eliminate one constant.
Using the expressions  in Appendix  one concludes that under (\ref{resc})
the integrals scale as
%Let us see how the integrals scale under rescaling (\ref{resc})
\begin{equation}
I_1 \rightarrow \frac{1}{\epsilon^2}I_1, \quad \quad I_2 \rightarrow \frac{1}{\epsilon^4}I_{+}, \quad \quad I_3 \rightarrow \frac{1}{\epsilon^4}I_{+},
\quad \quad I_5 \rightarrow I_4, \quad \quad I_6 \rightarrow I_4
\label{rescI}
\end{equation}
where $I_{+}= \frac{1}{2}(I_2+I_3)$.
% and $I_4$ is defined in appendix A.
The constraint (\ref{cak}) that determines  $\beta$
%under this rescaling
becomes
\begin{equation}
2 c_0 I_4 + w_0 \beta I_1=0  \label{qfk}
\end{equation}
The equation for $v$  \rf{vew} retains its  form in this limit
\begin{equation}
v'= \frac{\sqrt{2 v P(v)}}{1-\beta^2}, \qquad \qquad P(v)= - 8 c_0^2 (v-v_1)(v-v_2)(v-v_3)
\end{equation}
%where $\beta$ which is finite can be replaced from (\ref{qfk}).
The cubic polynomial $P(v)$ (\ref{poli})  can be written also
as
\begin{equation}
P(v)=v^3 (-4 C_0^2 -4 C_1^2)+v^2 (4 C_0^2-4 C_1^2) +v [w_0^2+w_1^2-2 (D^2+\nu^2)] +w_0^2-w_1^2
\end{equation}
%To further extract the proper leading order polynomial we
 Introducing the parameters
\begin{equation}
w_{\pm}= \frac{w_1 \pm w_0}{2}, \qquad \qquad\qquad  C_{\mp}= \frac{C_1 \pm C_0}{2}
\end{equation}
%Observing that in the $pp$-wave the above parameters scale as
we find that they scale as
\begin{equation}
C_{-} \sim O(\epsilon^0), \quad \quad C_{+} \sim \frac{1}{\epsilon^2}, \quad \quad w_{-} \sim \epsilon^2, \quad \quad w_{+} \sim O(\epsilon^0)
\end{equation}
%we obtain the following polynomial in the limit
Then the scaling limit of $P(v)$ takes the form
\begin{equation}
P(v)= - 2 v^3 C_{+}^2 -4 C_{-} C_{+} v^2 + 2 (w_{+}^2 - \nu^2 -D^2) v - 4 w_{+} w_{-}  \label{tiu}
\end{equation}
%Let us compute the charges after rescaling
The number of spikes $n$  and the winding number $m$   are given by
\begin{equation}
{n}=    \frac{\pi w_0^2}{2c_0^2 v_1 v_2 v_3 I_4}\ {1 \ov \epsilon^2},\qquad \qquad
 \frac{m}{n}=- \frac{2 c_0 \nu I_4+ D w_0 I_1}{4 \pi  c_0 w_0}
\end{equation}
We see that  in this
scaling  limit $n$ grows to infinity  while $\frac{m}{n}$ stays finite.

The scaling limit (\ref{resc})  leads also to the following expressions for the conserved charges
\begin{equation}
\frac{\mathcal{E}}{4\pi n}= \bigg(\frac{w_0}{8 c_0}I_{+} - \frac{c_0}{2 w_0}I_4\bigg)\frac{1}{\epsilon^2},
\quad \quad \frac{\mathcal{S}}{ 4\pi n}= \bigg(\frac{w_1}{8 c_0}I_{+}- \frac{c_0}{2 w_0}I_4\bigg)\frac{1}{\epsilon^2}, \quad \quad
\frac{\mathcal{J}}{4\pi n}= \frac{w_0 \nu I_1 +2 D c_0 I_4}{4 c_0 w_0}
\nonumber
\end{equation}
Thus  the ratio $\frac{\mathcal{J}}{n}$ remains finite while  $\frac{\mathcal{E}}{n}$ and
 $\frac{\mathcal{S}}{n}$ diverge in this
 limit.

We then find also that
\begin{equation}
\frac{\mathcal{E}-\mathcal{S}}{2n}= \frac{c_0}{2 w_0}v_1 v_2 v_3 I_{+}, \qquad \qquad \frac{\mathcal{E}+\mathcal{S}}{2n}=
\bigg(\frac{w_0+w_1}{8 c_0}I_{+} - \frac{c_0}{w_0}I_4\bigg) \frac{1}{\epsilon^2} \sim n
\end{equation}
%which are as we expect. We expect $\frac{\mathcal{E}-\mathcal{S}}{2n}$ to be finite while $\frac{\mathcal{E}+\mathcal{S}}{2n}$ to scale as $\frac{1}{\epsilon^2}$, which is indeed the case.
%Multiplying charges by $T$ we observe that this limit corresponds to $\frac{J}{n}$ fixed but large. We also have that
We conclude that in this limit
\begin{equation}
\mathcal{S} \gg \mathcal{J} \sim n \gg 1
\end{equation}
%Thus in the $pp$-wave limit the charges scale as
and
\begin{equation}
\frac{\mathcal{E}+\mathcal{S}}{n^2} \sim \frac{\mathcal{E}-\mathcal{S}}{n} \sim \frac{\J}{n}
\sim \frac{m}{n}\ \  =  \ \   \texttt{finite}
\end{equation}
Let us note that a property of the scaling
limit we discussed above is that
 the number of independent parameters gets reduced
  by one. Namely, in this  $pp$-wave limit there remain
  only three independent parameters.
   To see this let us  perform  the following rescaling
\begin{equation}
v_i \rightarrow c v_i, \quad \quad C_{+}\rightarrow c^{-1} {C_{+}},
 \quad \quad w_{-} \rightarrow  c  w_{-}
\end{equation}
with  all other parameters kept  fixed. Under this
rescaling the charges and the polynomial (\ref{tiu}) remain unchanged,
 so that  all the
 quantities depend only on the combination $C_{+} w_{-}$ and not separately on
  $C_{+}$ and $w_{-}$.

\section{Fast spiky string from  $SL(2,R)$ Landau-Lifshitz  model}

As shown in \cite{st,ptt} following   \ci{k,krt2},
 taking a large $S^5$ orbital momentum   limit of the classical
string action one can truncate it to a non-relativistic Landau-Lifshitz (LL) action
that should thus be describing  the
 fast-moving string  solutions to leading order in expansion in
large $\J$.  The same  LL action happens to arise also
 from the  1-loop  dilatation operator
in the $SL(2)$ sector of the SYM  theory  when  one restricts  consideration
to certain  ``locally-BPS'' coherent states present in the thermodynamic limit of large spin chain
length $J$.
In this  limit (which is a generalisation of the  BMN limit \ci{bmn})
 the one-loop  gauge theory result for the energy happens  to match
the leading-order term  in the fast-string  energy  \ci{ft2,mik,ft4,bfst}.

%In this case the string side matches a one-loop result from the field theory side as in \cite{bmn}.
The leading-order fast-string expressions  obtained in the previous section should  thus follow also
 from a particular solution of the LL model  \cite{ptt}\foot{Here we made a rescaling by a factor of
$\J= {J\ov \sql}$
 which will be restored later  in the expressions for conserved charges.}
\begin{equation}
 L = \dot{\eta} \sinh^2 \rho
- \frac{1}{2 } \Big( \rho'^2 + \ha  \sinh 2\rho\  \eta'^2  \Big)
\label{eqn:landau}
\end{equation}
We shall use the ansatz
\begin{equation}
  \eta = \omega \tau + f(\sigma), \qquad \qquad\rho = \rho(\sigma)
\end{equation}
%Then, the equations of motions for $\eta$ and $\rho$ are, respectively,
leading to
\begin{equation}
  \partial_\sigma \left(f' \sinh^2\rho \cosh^2 \rho \right) =0  \ , \qquad \qquad
  \rho'' = \frac{1}{2} \sinh 2\rho\  \left( f'^2 \cosh 2 \rho - 2 \omega  \right)
\label{e:rho}
\end{equation}
%Since the inside of the derivative is constant, these equations lead,
Then
\begin{equation}
  f' = \frac{4 C_2}{\sinh^2 2 \rho}, \qquad\qquad
  \rho' = \sqrt{2 C_3 - \omega  \cosh 2\rho - 4C_2^2 \coth^2 2\rho}
\end{equation}
where $C_2, C_3$ are integration  constants. Equivalently,  the equation for $\rho$   can be  written as
\bea
 && v' = \sqrt{2 v P(v)} \ , \ \ \ \ \ \ \ \ \ \ \ \ \ \ v={1\ov \cosh 2\rho}
  \label{eqn:pp}
  \\
&& P(v) = -4 C_3 v^3 + 2 \omega v^2 + 4(C_3-2 C_2^2) v - 2 \omega
= -4 C_3 (v-v_1) (v-v_2) (v-v_3)
\ee
Now we can follow the same argument as in the previous section.
Namely, we assume that $P(v)$ has two positive roots $0<v_2<v_3$, and
$v_1<0$. The product of the roots is related to
\begin{equation}
 b \equiv - 2 C_3 = \frac{\omega }{v_1v_2v_3}
\end{equation}
Again,  we have $b<0$,
so that  for a  physical solution  $v_2 < v \leq v_3$.
One can find $v_1, C_2, C_3$ in terms of $v_2, v_3$
\begin{equation}
 C_2^2 = -\frac{\omega (1-v_2^2)(1-v_3^2)}{4v_1v_2v_3(1+v_2v_3)},
 \qquad C_3 = -\frac{\omega }{2v_1v_2v_3},
 \qquad v_1= - \frac{v_2+v_3}{1+v_2v_3}
\end{equation}
The number of spikes $n$ is determined by
\begin{equation}
  2 \pi = \Delta \sigma = \int_0^{2\pi}d\sigma
        = \int \frac{d\rho}{\rho'}
        = 2 n \int_{v_2}^{v_3} dv \frac{d\rho}{dv}\frac{1}{\sqrt{vP(v)}}
        = \frac{n}{\sqrt{-b}} I_1
  \label{e:dels}
\end{equation}
where $I_1$ is given in (\ref{eqn:elliptic}).
The spin and energy of this solution  are given by
\begin{eqnarray}
 S &=&J \int_0^{2\pi} \frac{d \sigma}{2 \pi} \frac{\partial L}{\partial\dot{\eta}}
  = \frac{2 n}{2 \pi} J \int_{v_2}^{v_3} dv \frac{d\rho}{dv}\frac{\partial L}{\partial\dot{\eta}}
  = \frac{n I_2}{4 \pi \sqrt{-b}}\ J   \\
 E-S-J &=&\frac{\lambda}{J} \int_0^{2\pi} \frac{d \sigma}{2 \pi}
       \Big( \dot{\eta}\frac{\partial L}{\partial\dot{\eta}}
         + \dot{\rho}\frac{\partial L}{\partial\dot{\rho}} - L \Big)
   =  \frac{2 n}{2 \pi} \frac{\lambda}{J} \int_{v_2}^{v_3} dv \frac{d\rho}{dv}
        \Big( \dot{\eta}\frac{\partial L}{\partial\dot{\eta}}
        + \dot{\rho}\frac{\partial L}{\partial\dot{\rho}} - L \Big) \notag \\
   &=& \frac{n \lambda
}{4 \pi J }\
         \sqrt{-b}
\Big[v_1v_2v_3 I_+ - (v_1v_2+v_2v_3+v_3v_1)I_1 \Big]
\end{eqnarray}
where $I_+=\frac{1}{2}(I_2+I_3)$, and  $I_1, I_2, I_3$ are again
defined  in (\ref{eqn:elliptic}).

The winding number is determined from an  additional constraint given in \cite{ptt}
\begin{eqnarray}
 m &=& \int_{0}^{2\pi}\frac{d\sigma}{2\pi} \  i V^*_r \del_\sigma V^r
        +  O(\frac{1}{\mathcal{J}^2})  \nonumber  \\
   &=& {n\ov \pi} \int_{v_2}^{v_3} dv \frac{d\rho}{dv}\frac{1}{\sqrt{vP(v)}}
           \  i V^*_r \del_\sigma V^r  + O(\frac{1}{\mathcal{J}^2}) \nonumber \\
   &=& {n\ov 2 \pi}  \sqrt{1+ v_1v_2+v_2v_3+v_3v_1} I_5
 \label{constraint}
\end{eqnarray}
where $V^*_r V^r= V_0^2 - |V_1|^2=1$  with
    $V_0 = \cosh \rho$,\  $V_1 = \sinh \rho\  e^{i \eta}$.

Eliminating $\sqrt{-b}$ using  (\ref{e:dels}) we thus end up with
\begin{eqnarray}
  \frac{S}{J} &=& \frac{I_2}{2 I_1}\ , \\
  E -S -J &=& \frac{n^2 \lambda}{8 \pi^2J}\
         I_1 \Big[v_1v_2v_3 I_+ - (v_1v_2+v_2v_3+v_3v_1)I_1 \Big]\ , \\
  m  &=& { n\ov 2 \pi}  \sqrt{1+v_1v_2+v_2v_3+v_3v_1}\ I_5
  \label{e:sum}
\end{eqnarray}
Below we shall look at several special cases of these expressions.

\subsection{Folded string:  $n=2$}

The 2-spike case corresponds to
$$ v_1=-1\ ,    \ \ \     \ \ \ \ \ \ \      v_3=1 \ , \ \ \ \  v_2={\rm arbitrary}   $$
%with  $v_2$ arbitrary.
 Then
\begin{eqnarray}
  \frac{S}{J} &=& \frac{1}{2} \Big(1+\frac{1}{v_2} \Big)
   \frac{\eE[ q]}{\eK[q]}-1 , \ \ \ \ \ \ \ \ \ \ \       q\equiv \sqrt{\frac{1-v_2}{1+v_2}}\\
  E-S-J &=&\frac{4\lambda}{2 \pi^2J}\  \eK[q]
          \left(  \eK[q] - \eE[q] \right) , \\
  m  &=& 0
\label{e:n2}
\end{eqnarray}
Here
\be  v_2={1 \ov \cosh 2\rho_2} = {1 \ov 1 - 2 x_0} \ , \ \ \ \ \ \ \
x_0\equiv  - \sinh^2 \rho_2 \ee
so that   we can transform the elliptic functions as\foot{We use that \cite{Grad}: \  \
$\eK[\frac{ik}{k'}]=k' \eK[k], \quad  \eE[\frac{ik}{k'}]=\frac{\eE[k]}{k'}
 $ {where}\ $ k'=\sqrt{1-k^2}$.}
\begin{equation}
  \eK[\sqrt{x_0}] = \frac{1}{\sqrt{1-x_0}} \eK[q],
  \qquad\qquad
  \eE[\sqrt{x_0}] = \sqrt{1-x_0}\eE[q]
\end{equation}
%where $x_0= - \sinh^2 \rho_2$, and we used the relation given in \cite{Grad},
%\begin{equation}
% K[\frac{ik}{k'}]=k' K[k], \qquad E[\frac{ik}{k'}]=\frac{E[k]}{k'}
% \qquad \text{where}\quad k'=\sqrt{1-k^2}
%\end{equation}
Then we finish with
\begin{eqnarray}
  \frac{S}{J} &=& \frac{\eE[\sqrt{x_0}]}{\eK[\sqrt{x_0}]} - 1, \\
  E-S-J &=&-\frac{2\lambda}{\pi^2J} \  \eK[\sqrt{x_0}]
          \Big(  \eE[\sqrt{x_0}] - (1-x_0) \eK[\sqrt{x_0}] \Big)
\end{eqnarray}
which are  the same as equations (B.18) and (B.19)  in
\cite{bfst}.\footnote{Note that ref. \cite{bfst} used a different  definition of elliptic
functions:  here  $\eK[\sqrt{x_0}]$ is the same as  $\eK(x_0)$
 in \cite{bfst}, etc. }

\subsection{Near-boundary, fixed number of spikes: ``long-string''  limit}

Next, let us look at another particular limit:
$$v_1=-v_3,\qquad \qquad  v_2\rightarrow 0 $$
In this case
\begin{eqnarray}
  \frac{S}{J} &\simeq& \frac{1}{v_2 \ln \frac{8v_3}{v_2}}+O(1) , \\
  E -S -J &\simeq& \frac{n^2 \lambda}{8 \pi^2J}\  \ln^2 v_2 + O(\ln v_2)
        , \\
  m  &\simeq& { n\ov 2 \pi }  \arccos v_3
\end{eqnarray}
where we used that for  $v_1=-(v_2+v_3)/(1+v_2v_3)$ one has
\begin{equation}
  I_5 \simeq \frac{\arccos v_3}{\sqrt{1-x_3^2}} \quad
  \text{as} \quad v_2 \rightarrow 0
\end{equation}
Solving for  $v_2$  we finish with
\begin{equation}
  E - S - J \simeq \frac{ n^2 \lambda}{8 \pi^2J} \  \ln^2 \frac{S}{J}
 \ + \ O\Big(\ln \frac{S}{J} \Big) \ .
\end{equation}
For $n=2$  this  reproduces eq.(1.5) in \cite{ftt}
for the corresponding  asymptotics of the energy of  the long fast-moving
 folded  string.
% when $n=2$.

\subsection{
%$pp$-wave limit:
 $n \to \infty$ limit }

To obtain the solution in the analog of the
$pp$-wave limit let us  perform  the same rescaling
(\ref{resc}) as in the previous section.
Again, $\omega$ is fixed while
\begin{equation}
  b \rightarrow \frac{1}{\epsilon^6} b,
  \qquad C_2 \rightarrow \frac{1}{\epsilon^{3}} C_2,
  \qquad C_3 \rightarrow \frac{1}{\epsilon^6} C_3,
  \qquad v_1 = - (v_2+v_3)
\end{equation}
The number of the spikes (\ref{e:dels}) then  scales as  (cf.(\ref{rescI}))
\begin{equation}
  n \rightarrow \frac{1}{\epsilon} n
\end{equation}
%Therefore, under this rescaling, the $\bar{m}$ and $\bar{\Delta}$ become
As a result, we find
\begin{equation}
 \bar m\equiv    \frac{m}{n} = {I_4\ov 2 \pi} \  ,  \label{puq}
\end{equation}
\begin{equation}
  \bar{\gamma} \equiv  \frac{E-S-J}{n} =  \frac{\lambda}{8 \pi^2\bar{J}}  \  {I_1}
        \Big[v_1v_2v_3 I_+ - (v_1v_2+v_2v_3+v_3v_1)I_1 \Big]  \label{puq1}
\end{equation}
where $\bar{J}=J/n$, and we used (\ref{rescI}) and that
$I_+\rightarrow \frac{1}{\epsilon^4} I_+$.
We see that $\bar{m}$ and $\bar{\gamma}$ are invariant under the   rescaling
$
  v_2 \rightarrow   c   v_2,  \ \  v_3 \rightarrow  c  v_3,
$
 i.e. they depend only on $v_2/v_3$.
 This means we can express $\bar \gamma$ as
 \begin{equation}\la{gah}
 \bar \gamma = \frac{\lambda}{8 \pi^2 \bar{J}}\ f(\bar{m})
 \end{equation}
where $f(\bar{m})$ can,  in principle,  be  computed
 from  (\ref{puq}),(\ref{puq1}).

 Since  the LL  model describes  also a  certain class of coherent  gauge-theory
  states in  the thermodynamic limit of the 1-loop gauge-theory
 $SL(2)$ spin   chain,  the above  expression  should  represent  the  1-loop
  anomalous dimension of the corresponding    ``long'' ``locally-BPS''
  gauge-theory operator.\foot{Let us note that the large $n$ limit
considered in this subsection is not exactly the same as
the pp-wave limit considered above. The assumption made
  in deriving the LL model is that $J \gg 1$  with
$S\ov J$ (and $m$) kept fixed, while in the pp-wave limit
we had $S\ov J$ $ \sim n \gg 1$.  In general, the  fast-string or
 LL limit and the  pp-wave limit do not commute.  As a consequence,
we cannot set $m=0$ in the above expression \rf{gah}.}

% Thus,
%we can solve $\bar{m}$ about $v_2/v_3$ and plug in $\bar{\Delta}$.  As a
%consequence, this determines $\bar{\Delta} = \bar{\Delta}(\bar{J},\bar{m})$. In this case the
% result is valid for both,
%the string side and the field theory side since the LL model can be thought as describing both.

\section{Periodic spikes in AdS--pp-wave $\times$  $S^1$  background }

Let us now discuss a generalization of the  periodic  spike solution from
section 2 to the case of non-trivial  motion in $S^5$.

After  taking the limit  discussed in  \cite{ppwave}
the metric  becomes an AdS--pp-wave times original
$S^5$
 (the limit did not affect the 5-sphere).
Let us  consider  the solution in the following
4-dimensional    subspace of this limiting space   with metric
%($x_i=0$)
\beq
 ds^2 = \frac{1}{z^2} \left(dz^2 + 2 dx_+ dx_- - \mu^2 z^2 dx_+^2 \right) + d\alpha^2
\eeq
where $\alpha$ is an angle of period $2\phi$ parameterizing a maximal circle
 $S^1\subset S^5$.
 It is straightforward
 to write down the equations of motion and
constraints in  conformal gauge for a string moving in such background.
 To solve them, we propose, by analogy with  the discussion in the previous sections,
  the following  ansatz
\beq
x_\pm = \omega_\pm \tau + \phi_\pm(u) , \ \  \ \ \alpha=\omega \tau + \phi_\alpha(u),\ \  \ \ \ z=z(u),
\  \ \ \ \ \ \ \  u\equiv \sigma+\beta\tau\label{ppan}
\eeq
%where $u=\sigma+\beta\tau$.
 Such solution represents a rigid string that moves  along the direction $x= \sqrt 2 (x_+ + x_-)$.
 In addition, it wraps the $S^1$ as well as
moves along it. The equations of motion for $x_\pm$ and $\alpha$ can now be easily integrated to
\beq
\phi'_\alpha = C_{\alpha}  , \ \ \ \phi'_+ = \frac{1}{1-\beta^2}\left( \beta\omega_+ + C_+ z^2\right),
\ \ \ \phi'_- = \frac{1}{1-\beta^2}\left(\beta\omega_-+C_- z^2 + \mu^2C_+ z^4\right)
\eeq
where $C_\alpha$, $C_{\pm}$ are constants of integration. The conformal gauge  constraints are
\beqa
0&=& C_-\omega_+ + C_+\omega_-+ \omega D  \\
(1-\beta^2)^2z'{}^2 &=& -\mu^2 C_+^2 z^6 - 2 C_+ C_- z^4 + (\mu^2 \omega_+^2 - \omega^3 - D^2)z^2 -2\omega_+\omega_-
\eeqa
where, for simplicity, we introduced a new constant $D$ through
\beq
C_\alpha = \frac{\beta \omega -D}{1-\beta^2}
\eeq
Introducing the  variable $v=z^2$ we can rewrite the second constraint as
\beqa
 v' &=& \frac{\sqrt{2vP(v)}}{1-\beta^2}  \\
 P(v) &=& -2\mu^2 C_+^2 v^3 - 4C_+C_-v^2 + 2(\mu^2\omega_+^2 - \omega^2-D^2) v - 4\omega_+\omega_-
\eeqa
 This equation  determines the shape of the string $z(u)$  (given   which
 we can then obtain $x_{\pm}(\sigma,\tau)$).
 The same equation   can be found
by taking the limit discussed in section 5C  (cf. eq. (\ref{tiu})).
 Here we  rederived this  result  as a check and also to introduce the ansatz (\ref{ppan})
which can be useful to obtain other solutions.

\section{Conclusions}

In this paper we have analyzed
%, in conformal gauge,
 a large class of string solutions in $AdS_3 \times S^1$.
 We focused in  particular
on the spiky string solutions in $AdS_3 \times S^1$ and their
limits.
%large $n$ or ``$pp$-wave'' limits.
We found that because of the motion in $S^1 \subset S^5$ the
spikes no longer   end on cusps.

In the limit of fast spiky string we matched
 its  energy to that of the corresponding solution
in the $SL(2,R)$ Landau-Lifshitz   model.

Another limit that we considered was  the ``$pp$-wave'' limit in which
the number of spikes $n$
goes to infinity  while other conserved quantities scale in certain ways with $n$.
% while $\bPp=\frac{\mathcal{E}+\mathcal{S}}{n^2}$,
 %  $\bar{\Delta}=\frac{\mathcal{E}-\mathcal{S}}{n}$,   $\bar{J}=\frac{J}{n}$ and
% $\bar{m}=\frac{m}{n}$ stay finite. Here $E,S$ are the energy and spin in $AdS_3$ and $J,m$ are the
%angular momentum and winding number in $S^1$.

On the gauge theory side this limit corresponds
to a particular thermodynamic limit of the $SL(2)$ spin chain.
The classical  string   energy that we found  represents a strong-coupling prediction
for the corresponding gauge theory anomalous dimension.
The  implications of this limit   deserve further study.

%In that limit we then make a strong coupling prediction for the
%relation between
%$\bar{\Delta}$, $\bPp$, $\bar{J}$ and $\bar{m}$.

\bigskip

\section*{Acknowledgments }
%%%%%%%%%%%%%%%%%%%%%%%%%%
We are  grateful to P. Argyres, N. Dorey and L. Pando Zayas for useful discussions.
 M.K. and A.T. were supported in part by NSF under grant PHY-0847322.
The work of R.I. was supported in part by the Purdue Research Foundation.

\appendix
%\addcontentsline{toc}{section}{Appendices}
%\addcontentsline{toc}{section}{Appendices}
\subsection*{Appendix:  Some useful integrals
}
Here we summarize
various integrals that we used  above
\be
I_1= \int_{v_2}^{v_3} d v \frac{1}{\sqrt{-v (v-v_1)(v-v_2)(v-v_3)}},
\quad I_2= \int_{v_2}^{v_3} d v\frac{1-v}{v\sqrt{-v (v-v_1)(v-v_2)(v-v_3)}}
\nonumber
\ee
\begin{equation}
I_3= \int_{v_2}^{v_3} d v\frac{1+v}{v\sqrt{-v (v-v_1)(v-v_2)(v-v_3)}}
\end{equation}
These integrals can be written in terms of the elliptic integrals
\begin{equation}
I_1= \frac{2}{\sqrt{v_3 (v_2-v_1)}}\eK[\sqrt{s}], \quad I_2= -\frac{2}{v_1 v_2}\sqrt{\frac{v_2-v_1}{v_3}} \eE [\sqrt{s}]
+\frac{2 (\frac{1}{v_1}-1)}{\sqrt{v_3 (v_2-v_1)}}\eK[\sqrt{s}]
\label{eqn:elliptic}
\end{equation}
\begin{equation}
I_3= -\frac{2}{v_1 v_2}\sqrt{\frac{v_2-v_1}{v_3}} \eE [\sqrt{s}]+ \frac{2 (\frac{1}{v_1}+1)}{\sqrt{v_3 (v_2-v_1)}}\eK[\sqrt{s}]
\end{equation}
where
\begin{equation}
s= \frac{v_1 (v_2-v_3)}{v_3 (v_2-v_1)}
\end{equation}

Other integrals we used in this paper are
\begin{eqnarray}
I_4(v_2,v) &=&  \int_{v_2}^{v } d v \frac{v}{\sqrt{-v (v-v_1)(v-v_2)(v-v_3)}} \\
I_5 (v_2,v)&=& \int_{v_2}^{v} d v \frac{v}{(1+v) \sqrt{-v (v-v_1)(v-v_2)(v-v_3)}} \\
I_6 (v_2,v) &=& \int_{v_2}^{v} d v \frac{v}{(1-v) \sqrt{-v (v-v_1)(v-v_2)(v-v_3)}}
\end{eqnarray}

For particular $v=v_3$ these integrals can be written as
\begin{eqnarray}
I_4 \equiv I_4(v_2,v_3) &=& \int_{v_2}^{v_3} d v \frac{v}{\sqrt{-v (v-v_1)(v-v_2)(v-v_3)}} \\
  &=& \frac{2 v_2}{\sqrt{v_3(v_2-v_1)}}\Pi [\frac{v_3-v_2}{v_3},\sqrt{s}] \nonumber \\
I_5 \equiv I_5(v_2,v_3) &=& \int_{v_2}^{v_3} d v \frac{v}{(1+v) \sqrt{-v (v-v_1)(v-v_2)(v-v_3)}} \\
  &=& \frac{2 v_2}{(1+v_2)\sqrt{v_3(v_2-v_1)}}\Pi [\frac{v_3-v_2}{v_3(1+v_2)},\sqrt{s}] \nonumber \\
I_6 \equiv I_6(v_2,v_3) &=& \int_{v_2}^{v_3} d v \frac{v}{(1-v) \sqrt{-v (v-v_1)(v-v_2)(v-v_3)}} \\
  &=& \frac{2 v_2}{(1-v_2)\sqrt{v_3(v_2-v_1)}}\Pi [\frac{v_3-v_2}{v_3(1-v_2)},\sqrt{s}] \nonumber
\end{eqnarray}

%The last integrals can be equivalently expressed as
%\begin{eqnarray}
%I_4 &=& \int_{v_2}^{v_3} d v \frac{1}{(1+v) \sqrt{-v (v-v_1)(v-v_2)(v-v_3)}}= \frac{2}{(1+v_1) \sqrt{v_3 (v_2-v_1)}}K[\sqrt{s}]\nonumber\\
%&-& \frac{2 (v_3-v_1)}{(1+v_3) (1+v_1) \sqrt{v_3 (v_2-v_1)}}\Pi[\frac{(v_2-v_3)(1+v_1)}{(v_2-v_1)(1+v_3)},\sqrt{s}]
%\end{eqnarray}
%\begin{eqnarray}
%I_5&=& \int_{v_2}^{v_3} d v \frac{v}{(1+v) \sqrt{-v (v-v_1)(v-v_2)(v-v_3)}}= \frac{2 v_1}{(1+v_1) \sqrt{v_3 (v_2-v_1)}}K[\sqrt{s}]\nonumber\\
%&+& \frac{2 (v_3-v_1)}{(1+v_3)(1+v_1) \sqrt{v_3 (v_2-v_1)}}\Pi[\frac{(v_2-v_3)(v_1+1)}{(v_2-v_1)(v_3+1)},\sqrt{s}]
%\end{eqnarray}
%\begin{eqnarray}
%I_6&=& \int_{v_2}^{v_3} d v \frac{v}{(1-v) \sqrt{-v (v-v_1)(v-v_2)(v-v_3)}}= \frac{2 v_1}{(1-v_1)\sqrt{v_3 (v_2-v_1)}}K[\sqrt{s}]\nonumber\\
%&+& \frac{2 (v_3-v_1)}{(1-v_3)(1-v_1)\sqrt{v_3 (v_2-v_1)}}\Pi[\frac{(v_2-v_3)(1-v_1)}{(v_2-v_1)(1-v_3)},\sqrt{s}]
%\end{eqnarray}

%%%%%%%%%%%%%%%%%%%%%%%%%%%%%%%%%%%%
%\begin{thebibliography}{20}
%%%%%%%%%%%%%%%%%%%%%%%%%%%%%%%%%%%%%%%%%%%%%%%%%%%%%%%%%

\newpage

%{\bf REFERENCES}

\end{document}